\documentclass{iopart}
\usepackage{iopams}
\usepackage{amssymb}
\usepackage{graphicx,subfigure,epsfig}
\usepackage{cite}  
\newcommand{\group}[1]{\mathsf{#1}}
\begin{document}
\title{Analytic properties of the free energy: the tricritical Ising model}
\author{Alessandro Mossa \footnote{Present address: Departament de F\'{\i}sica Fonamental, Facultat
de F\'{\i}sica, Universitat de Barcelona, Diagonal 647, E-08028 Barcelona, Spain} and Giuseppe Mussardo}
\address{SISSA and INFN, via Beirut 2-4, I-34014 Trieste, Italy}
\eads{\mailto{alessandro\_mossa@ub.edu}, \mailto{mussardo@sissa.it}}
\begin{abstract}
We investigate the tricritical Ising model in complex magnetic field in order to characterize the analytic 
structure of its free energy. By supplementing analytic methods with the truncation of conformal space technique we obtain nonperturbative data even if the field theories we consider are not integrable.  The existence of edge singularities analogous to the Yang--Lee points in the Ising field theory is confirmed. A surprising result, due to the conformal dimensions of the operators involved, is the appearance of two branching points which seems appealing to identify with a pair of complex conjugate spinodal singularities. 
\end{abstract}
 \submitto{Journal of Statistical Mechanics: Theory and Experiment}

\section{Introduction}

More than fifty years ago, Yang and Lee~\cite{PR87:404,PR87:410} showed the importance of understanding the analytic structure of the free energy in the theory of phase transitions. They proved that the thermodynamical equation of state is completely determined by the distribution of roots of the partition function (or, equivalently, singularities of the free energy). Their papers inspired several lines of further research, which we will not attempt to review here. We just note that, as far as the Ising model is concerned, the task of unveiling the analytic properties of the free energy has been recently completed by Fonseca and Alexander Zamolodchikov~\cite{JSP110:527}. They have been able to substantiate the identification of the Yang--Lee edge singularity with the spinodal point, thus unifying the high- and low-temperature descriptions.

One of the main tools of Fonseca and Zamolodchikov's analysis is the truncated free-fermion space approach, a technique that gives numerical access to the lowest energy levels of the Ising field theory defined on a cylinder. A similar analysis can, in principle, be performed to study the next-to-simplest unitary minimal model, namely the tricritical Ising model (in the following often referred to by the acronym TIM). Due to the universality principle, the insight gathered by working on this conformal field theory and its perturbations applies to all tricritical phenomena in two dimensional space,  characterized by the same $\mathbb{Z}_2$ symmetry of the order parameter.

In practice, however, the extension of Fonseca--Zamolodchikov's method is far from trivial because the tricritical Ising model possesses four primary fields that are relevant in the renormalization group sense, instead of two like in the Ising case. On technical ground, moreover, it is not possible to exploit the equivalence between the magnetically perturbed Ising theory and a theory of free fermions, so that one is forced to employ a less powerful approximation, the truncation of the conformal space of the minimal model $\mathcal{M}(5,4)$. Nonetheless, some picture can be outlined, and the resulting sketch constitutes the subject of this article. 

In order to make the paper reasonably self-contained, the main points of Fonseca and Zamolodchikov's work, as well as earlier progress on the analytic structure of the Ising model in complex magnetic field, are collected in the next subsection. \Sref{sec:TCS} briefly presents the truncation of conformal space (TCS for the sake of brevity) technique that we adopt, while the most relevant known facts about the tricritical Ising model are summarized in \sref{sec:TIM}. The results of our investigations are thoroughly reported in \sref{sec:FE} and then synthesized and commented in the final section.  

\subsection{The Ising model in complex magnetic field}
 
\begin{figure}
 \centering
 \mbox{\subfigure[]{\epsfig{figure=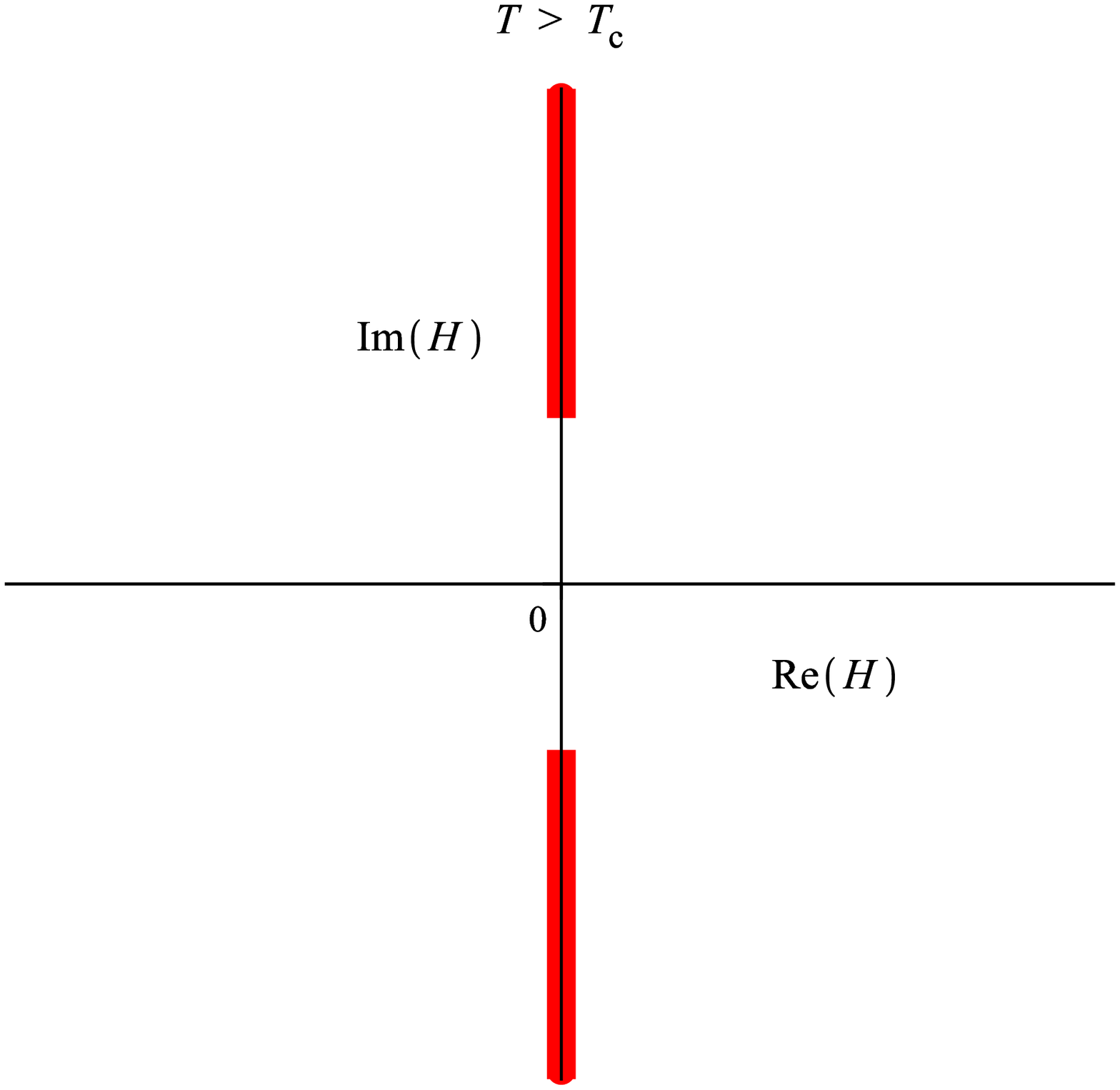,width=6cm \label{fig:HTIsing}}}\quad
             \subfigure[]{\epsfig{figure=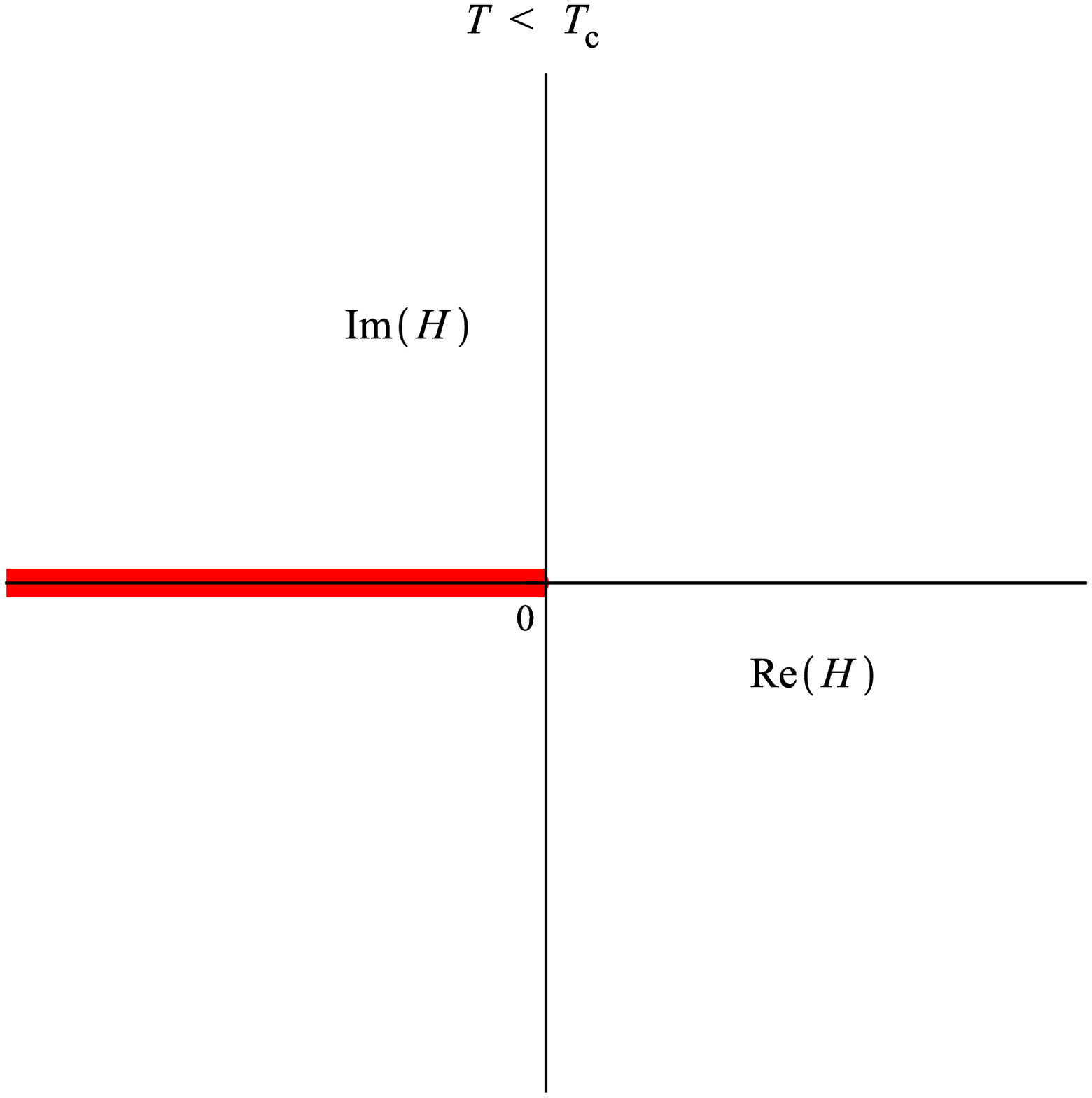,width=6cm \label{fig:LTIsing}}}}
 \caption[Ising model's free energy]{Ising model: branch cuts of the free energy in the complex $H$-plane. The magnetic field is measured in units of $|2\pi g_2|^{15/8}$. In the left panel, the high-temperature regime is considered and the Yang--Lee edge singularities are depicted. The right panel shows the Langer branch cut in the low-temperature regime.} 
\end{figure}

As written in the introduction, the study of the Ising model in a complex magnetic field $H$ was introduced by Yang and Lee in 1952. In their first paper~\cite{PR87:404} they studied the grand partition function of a lattice gas with complex fugacity $y$, proving that the equation of state can be deduced by the distribution of the singularities of the free energy $F$. The second paper~\cite{PR87:410} established the equivalence between the Ising model and a lattice gas with a suitable potential. In particular, the fugacity was found to be proportional to the exponential of the magnetic field:
\begin{eqnarray}
  y\propto \rme^{-2\beta H} \,.
\end{eqnarray}
Then a theorem was proved that, for a broad class of potentials (including the one corresponding to the Ising model), all the zeros of the grand partition function lie on the unit circle of the complex $y$-plane (equivalently, all the singularities of the free energy are located on the imaginary axis of the complex $H$-plane). 
 
For $\beta<\beta_\mathrm{c}$ (the high-temperature regime), the logarithmic singularities of the free energy accumulate in the thermodynamic limit towards the points $H=\pm \rmi H_0(\beta)$ with real $H_0>0$. To visualize the situation we can join these accumulation points by drawing a cut along the imaginary axis (see \fref{fig:HTIsing}). The cut passes through the point at infinity: the free energy is analytic for any real value of $H$ because we know there is no phase transition.

\begin{figure}
\begin{center}
 \includegraphics[width=6cm,height=6cm]{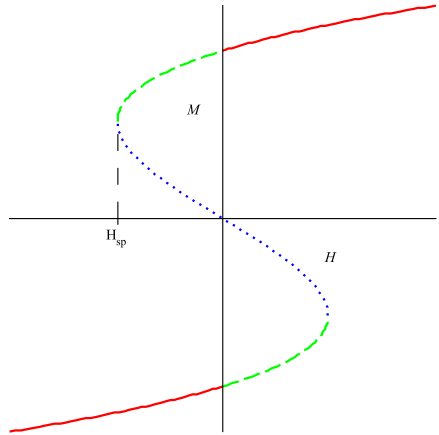}
 \caption[Spinodal point]{Magnetization $M$ versus the magnetic field $H$ for $\beta>\beta_\mathrm{c}$. The red solid line indicates the stable branches, the green dashed line are the metastable branches, while the blue dotted line represents the unstable branch. The value $H=H_\mathrm{sp}$ where the metastable branch turns into unstable is called spinodal singularity.} \label{fig:spinodal}
\end{center}
\end{figure}

The gap between the branch points reduces while lowering the temperature, until, at the critical temperature, it happens that $H_0(\beta_\mathrm{c})=0$. Now it is no more possible to analytically continue $F(\beta,H)$ from positive to negative $H$: this is the signal of the phase transition. Fisher~\cite{PRL40:1610} named Yang--Lee edge singularities the points $\pm \rmi H_0(\beta)$, and proved that the accumulation of zeros of the partition function in the thermodynamic limit is a conventional critical phenomenon, with scaling laws, universality and so on. In high dimension $D$, it corresponds to the infrared behaviour of the field theory with an action
\begin{eqnarray}
  \mathcal{A}_\mathrm{YL}=\int \rmd^Dx \left[ \frac{1}{2}(\partial_\mu\phi)^2+\rmi(h-h_0)\phi
  +\rmi\gamma\phi^3 \right] \,.
\end{eqnarray} 
Note that such a theory is nonunitary because of the imaginary couplings. The theory of Yang--Lee edge singularity was enriched by Cardy \cite{PRL54:1354} who showed how, at the 
critical point, it is related to the simplest nonunitary minimal model characterized by central charge $c=-22/5$. Away from the critical point, the Yang--Lee edge singularity is an integrable theory, whose exact $S$-matrix was proposed by Cardy and Mussardo \cite{PLB225:275} and its Form Factors and correlation functions were computed by Al.~Zamolodochikov \cite{NPB358:497}. 

On the other hand, for $\beta>\beta_\mathrm{c}$ (low-temperature regime), if one tries to continuously change the magnetic field from $H>0$ to $H<0$, at $H=0$ one enters a metastable phase (that is, the system is in a local free energy minimum). Eventually the spinodal point is reached, where the metastable phase becomes unstable (see \fref{fig:spinodal}). This point is a singularity, in the sense that (at least in a mean field approximation) the isothermal susceptibility diverges as one approaches it. In fact, the above picture must be corrected by taking into account thermic fluctuations \cite{AP41:108}. They make the system start decaying through nucleation before reaching the spinodal singularity. A cut (named Langer's branch cut) is therefore drawn along the negative $H$-axis (see \fref{fig:LTIsing}), starting from $H=0$ where is a weak singularity (see \cite{PRA29:341} for a mean field theory of spinodal points and first order phase transitions). 

What happens to the spinodal point when the Langer's branch cut is opened? Fonseca and Zamolodchikov answer that it is pushed under the cut in order to reappear in the high-temperature regime under the name of Yang--Lee edge singularity\footnote{It seems that the first to speculate about a possible connection between spinodal points and the Yang--Lee edge singularity was Klein in \cite{PRL47:1569}.}. This identification completes\footnote{In fact, the comparison with lattice results (see \cite{JSP102:795}) could still hide some tricky point.} the task of describing the analytical properties of the free energy in the complex magnetic field plane.

In order to support their claim, Fonseca and Zamolodchikov use the Ising field theory, formally defined by the action
\begin{eqnarray} \label{eq:Aift}
  \mathcal{A}_\mathrm{IFT}=\mathcal{A}_{(c=1/2)}+g_1\int\sigma(x)\rmd^2x+g_2\int\varepsilon(x)\rmd^2x \,.
\end{eqnarray}
In the above formula $\mathcal{A}_{(c=1/2)}$ represents the action of the simplest unitary minimal model $\mathcal{M}(4,3)$, $\sigma(x)$ and $\varepsilon(x)$ are the primary fields of the theory, that we can interpret as spin and thermal operators, while $g_1$ and $g_2$ are couplings related to departures from the critical point in the magnetic or in the temperature direction, respectively. Actually, it is possible to put together the first and the last terms on the RHS of \eref{eq:Aift} and substitute them with the free fermions action
\begin{eqnarray} \label{eq:Afft}
  \mathcal{A}_\mathrm{FF}=\frac{1}{2\pi}\int\left[\psi\bar{\partial}\psi
  +\bar{\psi}\partial\bar{\psi}+\rmi m\bar{\psi}\psi\right]\rmd^2x \,,
\end{eqnarray}
where $m=2\pi g_2$, and $\psi,\bar{\psi}$ are the two chiral components of a Majorana field. The operator $\sigma$ admits as well a representation in terms of the Majorana fermion and it is easy to compute its matrix elements -- at a finite volume -- on the multi-particle fermionic states \cite{JSP110:527,PLA319:390}.  
Whichever formulation one employs, when the Ising field theory is put on a cylinder, its radius $R$ acts as an infrared regulator, so that the spectrum is infrared finite. By means of a suitable truncation of the Hilbert space the problem of measuring the energy levels is reduced to the diagonalization of a finite dimensional Hamiltonian. One has access to the free energy by studying the infrared (that is, $R\to\infty$) behaviour of the ground state.

\section{Truncation of the conformal space} \label{sec:TCS}

\begin{figure} \begin{center}
  \epsfig{figure=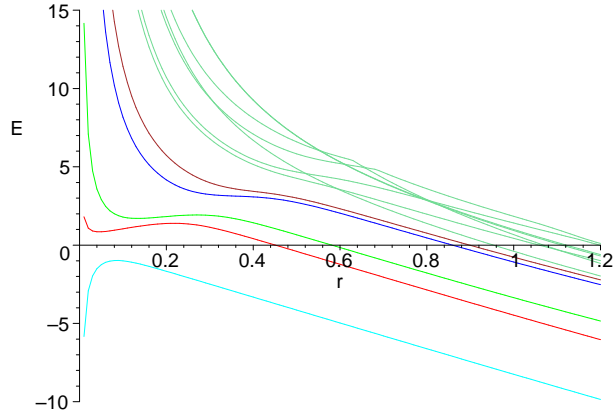, width=8cm}
  \caption[Spectrum of $\mathcal{A}_2^+$]{Spectrum of the theory $\mathcal{A}_2^+$: the first 12 levels.} \label{fig:tcsex1}
\end{center} \end{figure}

\begin{figure} \begin{center}
  \epsfig{figure=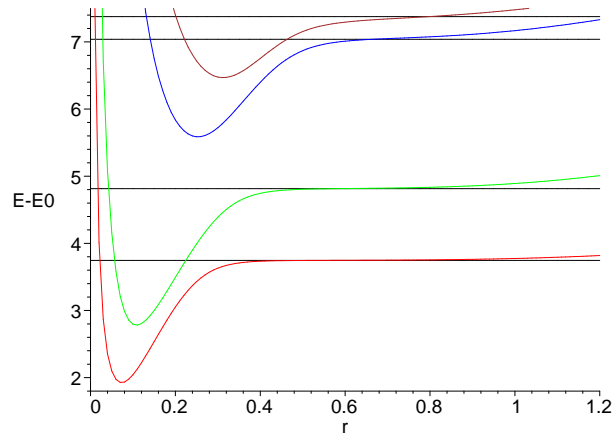, width=8cm}
  \caption[Masses of $\mathcal{A}_2^+$]{Spectrum of the theory $\mathcal{A}_2^+$: the black lines are the exact masses, the colored ones are energy levels differences $\Delta E_i=E_i-E_0$ as computed by TCS.} \label{fig:tcsex2}
\end{center} \end{figure}

The truncation of conformal space approach has been invented by Yurov and Alexei Zamolodchikov~\cite{IJMPA5:3221}. It is an approximate method that gives numerical access to the spectrum of two-dimensional off-critical theories arising from deformation of minimal models. Unlike other methods, however, the TCS works equally well for integrable and non-integrable deformations. 

The Hamiltonian acting on the Hilbert space is of course infinite dimensional; we need to truncate the conformal space if we want to be able to compute anything. In this paper the truncation is performed by discarding the states of level 6 or more in each Verma module. For the tricritical Ising model, this amounts to keep 228 states. The suitable truncated basis of the Hilbert space and the matrix elements of $\hat{H}_0$ and $\hat{V}_i$ have been computed by means of a program for Mathematica described in~\cite{CPC66:71}. The numerical diagonalization of the Hamiltonians was performed by the NAG routines of Maple.

This is a good place to say something about the precision of this approach. By following a standard procedure, introduced in \cite{NPB587:585} and recently theoretically justified \cite{preprintFeverati}, to keep under control the error produced by the truncation, in \cite{JSP110:527} the ground state energy for fixed $R$ was evaluated at different truncation levels $l$ and then extrapolated to $l\to\infty$ by fitting the formula
\begin{eqnarray}
  E_0^{(l)}(R)=s_0(R)+s_1(R)l^{-s_2(R)} \,.
\end{eqnarray}
However, the high precision of Fonseca and Zamolodchikov's numerical estimates is due above all to the fact that, dealing with the Ising model, they can exploit the free fermion basis, so that one of the two conformal perturbations is treated exactly (see \eref{eq:Afft}). This lucky accident is a peculiarity of the Ising model. Since we could not have reached such high precision anyway, we rather focused on the objective of gaining a qualitative understanding of the analytic structure of the free energy\footnote{The precision of the numerical data extracted by the truncation method can be improved by means of a renormalization group analysis, as shown  in \cite{preprintKonik}.}. 

A typical spectrum obtained by TCS is given in \fref{fig:tcsex1}. 
The first 12 levels of the theory $\mathcal{A}_2^+$ (defined in \sref{sec:offcrit}) are plotted against $r=R|g_2|^{9/5}/(2\pi)$. The lowest lines correspond to the ground state $E_0$ and the four lowest masses. Note how, above the threshold $E=2m_1$, we can find level crossings, signal of the integrability of the theory. In order to give an idea of the errors associated to this method, we can compare the exact masses of \tref{tab:e7} with the energy differences $\Delta E_i=E_i-E_0$, as in \fref{fig:tcsex2}.
It is quite evident that the TCS works better for the lowest masses. Another feature of TCS well exemplified by \fref{fig:tcsex2} is the existence of a physical window: we are interested in infrared data, that is in the limit $r\to\infty$, but as $r$ increases, the truncation effects become more and more relevant. 

The best precision, of course, is achieved when studying the ground state. In this case, the standard method to choose the physical window is to look at the effective scaling exponent of the ground state energy:
\begin{eqnarray}
  \alpha=\frac{R}{E}\frac{\rmd E}{\rmd R} \,.
\end{eqnarray}
The physical window is characterized by $\alpha\approx 1$. For excited levels, we found more convenient to look at the effective scaling exponent of the differences:
\begin{eqnarray}
  \alpha'=\frac{R}{E-E_0}\frac{\rmd E}{\rmd R} \,,
\end{eqnarray}
 selecting the physical window in the region where $\alpha'\approx 0$.

\section{The tricritical Ising model} \label{sec:TIM}

The class of universality of tricritical points (see \cite{PTCP9:1} for an extensive review) occurring in two-dimensional statistical models whose order parameter enjoys $\mathbb{Z}_2$ symmetry, for instance the Blume--Emery--Griffiths model \cite{PRA4:1071}, is described (in the scaling limit) by the minimal conformal field theory \cite{NPB241:333} characterized by central charge $c=7/10$. All data about this minimal model and its deformations relevant for understanding this paper are collected in this section.

\subsection{The minimal model $\mathcal{M}(5,4)$} \label{subsec:TIM}

The Kac table of the minimal model $\mathcal{M}(5,4)$ is given in \tref{tab:Kac}. 
\begin{table} 
 \caption{Kac table of the minimal model $\mathcal{M}(5,4)$.} \label{tab:Kac}
 \begin{indented}  
  \item[]\begin{tabular}{c c c c}
   \mr \ms 
    0 & $\frac{1}{10}$ & $\frac{3}{5}$ & $\frac{3}{2}$ \\[6pt]
   $\frac{7}{16}$ & $\frac{3}{80}$ & $\frac{3}{80}$ & $\frac{7}{16}$ \\[6pt]
   $\frac{3}{2}$ & $\frac{3}{5}$ & $\frac{1}{10}$ & 0 \\[6pt]
   \mr
  \end{tabular} 
 \end{indented}
\end{table} 
In the framework of radial quantization (see, for instance, \cite{CFT}), it is natural to define the theory on a cylinder by means of the conformal mapping $w=\frac{R}{2\pi}\ln\frac{z}{R}$. The operator content of the theory depends on the choice of boundary conditions \cite{NPB348:591} on the coordinate $u=\Re(w)$. 

For periodic boundary conditions, the modular invariant partition function is diagonal. The corresponding operator content is summarized in \tref{tab:TIM}: in the first column there are the conformal dimensions, the second and third columns report the symbols we will use in referring to these fields, while the fourth contains a concise description of their physical meaning. The rightmost column shows the correspondence \cite{SJNP44:529} with the normal ordered fields of a Landau--Ginzburg formulation. Fusion rules and structure constants of the operator algebra can be computed in the framework of the Coulomb gas formalism \cite{NPB240:312,NPB251:691}: the result is summarized in \tref{tab:struc}.

\begin{table} 
  \caption{Operators in the TIM with periodic boundary conditions.} \label{tab:TIM}
\begin{indented}  
\item[]\begin{tabular}{c l l l c}
     \br
     conf.~dim. & field & alias & phys.~role & LG field \\
     \mr 
    $(0,0)$ & $\mathbb{I}$ & & identity & \\[3pt]
    $(\frac{3}{80},\frac{3}{80})$ & $\sigma$ & $\varphi_1$ & magnetization & $\Phi$ \\[3pt]
    $(\frac{1}{10},\frac{1}{10})$ & $\varepsilon$ & $\varphi_2$ & energy & $:\Phi^2:$ \\[3pt]
    $(\frac{7}{16},\frac{7}{16})$ & $\sigma'$ & $\varphi_3$ & submagnetization & $:\Phi^3:$ \\[3pt]
    $(\frac{3}{5},\frac{3}{5})$ & $t$ & $\varphi_4$ & chemical potential & $:\Phi^4:$ \\[3pt]
    $(\frac{3}{2},\frac{3}{2})$ & $\varepsilon''$ & & (irrelevant) & $:\Phi^6:$ \\[3pt]
    \br
  \end{tabular} \end{indented}
\end{table} 

\begin{table} 
\caption[Fusion rules and structure constants for TIM]{Fusion rules and structure constants for TIM: periodic boundary conditions.} \label{tab:struc}
\begin{indented} 
\item[]\begin{tabular}{@{}l l}
\br
\centre{2}{even $\times$ even} \\
\mr
    $\varepsilon\times\varepsilon=[[\mathbb{I}]]+c\,[[t]]$ &  
    $t\times t=[[\mathbb{I}]]+c\,[[t]]$ \\[3pt]
    $\varepsilon\times t=c\,[[\varepsilon]]+\frac{3}{7}\,[[\varepsilon'']]$ & 
    $\varepsilon''\times\varepsilon''=[[\mathbb{I}]]$ \\
\mr
\centre{2}{even $\times$ odd} \\
\mr
    $\varepsilon\times\sigma=\frac{1}{2}\,[[\sigma']]+\frac{3}{2}c\,[[\sigma]]$ & 
    $\varepsilon\times\sigma'=\frac{1}{2}\,[[\sigma]]$ \\[6pt]
    $t\times\sigma=\frac{3}{4}\,[[\sigma']]+\frac{1}{4}c\,[[\sigma]]$ & 
    $t\times\sigma'=\frac{3}{4}\,[[\sigma]]$ \\
\mr
\centre{2}{odd $\times$ odd} \\
\mr
    $\sigma'\times\sigma'=[[\mathbb{I}]]+\frac{7}{8}\,[[\varepsilon'']]$ &
    $\sigma'\times\sigma=\frac{1}{2}\,[[\varepsilon]]+\frac{3}{4}\,[[t]]$ \\[3pt]
    \centre{2}{$\sigma\times\sigma=[[\mathbb{I}]]+\frac{3}{2}c\,[[\varepsilon]]+\frac{1}{4}c\,[[t]]+\frac{1}{56}\,[[\varepsilon'']]$} \\[9pt]
    \centre{2}{$c=\frac{2}{3}\sqrt{\frac{\Gamma(4/5)\Gamma^3(2/5)}{\Gamma(1/5)\Gamma^3(3/5)}}$} \\
    \br
\end{tabular}\end{indented}
\end{table}

If the boundary conditions on the $u$-direction are $\mathbb{Z}_2$-twisted, on the other hand, the partition function is not diagonal: the operator content of the theory is displayed in \tref{tab:twistTIM}, while the non-trivial fusion rules can be found in \tref{tab:twiststruc}. 

\begin{table} 
  \caption{Operators in the TIM with $\mathbb{Z}_2$-twisted boundary conditions.} \label{tab:twistTIM}
\begin{indented}  
\item[]\begin{tabular}{c l c}
     \br
     conf.~dim. & field & phys.~role \\
     \mr 
    $(\frac{3}{80},\frac{3}{80})$ & $\mu$ & disorder field \\[3pt]
    $(\frac{7}{16},\frac{7}{16})$ & $\mu'$ & subleading disorder field  \\[3pt]
    $(\frac{3}{5},\frac{1}{10})$ & $\psi$ &  fermion  \\[3pt]
    $(\frac{1}{10},\frac{3}{5})$ & $\bar{\psi}$ & anti-fermion  \\[3pt]
    $(\frac{3}{2},0)$ & $G$ & SuSy generator \\[3pt]
    $(0,\frac{3}{2})$ & $\bar{G}$ & SuSy generator \\[3pt]
    \br  
  \end{tabular} \end{indented}
\end{table} 

\begin{table} 
\caption[Fusion rules and structure constants for $\mathbb{Z}_2$-twisted TIM]{Fusion rules and structure constants for TIM: $\mathbb{Z}_2$-twisted boundary conditions.} \label{tab:twiststruc}
\begin{indented}    
\item[]\begin{tabular}{@{}l l}
\br
    $\psi\times\psi=\bar{\psi}\times\bar{\psi}=[[\mathbb{I}]]+c\,[[t]]$ &  
    $\psi\times\bar{\psi} =-\bar{\psi}\times\psi=\rmi c\, [[\varepsilon]]+\rmi \frac{3}{7}\,[[\varepsilon'']]$   \\[3pt]
    \centre{2}{$\psi\times G=-\bar{G}\times\bar{\psi}=\rmi \frac{3}{7}\,[[\varepsilon]]$} \\[9pt]
    \centre{2}{$c=\frac{2}{3}\sqrt{\frac{\Gamma(4/5)\Gamma^3(2/5)}{\Gamma(1/5)\Gamma^3(3/5)}}$} \\
    \br
  \end{tabular}
\end{indented}
\end{table}

The tricritical Ising model exhibits several discrete as well as continuous symmetries. First of all, there is the $\mathbb{Z}_2$ symmetry related to the spin-reversal transformation, that in the Landau--Ginzburg approach corresponds to $\Phi\to-\Phi$. The fields $\mathbb{I},\varepsilon,t,\varepsilon''$ are even with respect to such a transformation, while $\sigma,\sigma'$ are odd. An inspection of the fusion rules shows that the even operators form a subalgebra. 

Another symmetry mutuated from the lattice model is the Kramers--Wannier duality \cite{PR60:252,PR60:263}, under which the magnetization operators $\sigma,\sigma'$ are mapped onto their corresponding disorder operators $\mu,\mu'$, while $\varepsilon,\varepsilon''$ are odd, and $t$ is even. The behaviour of primary operators of TIM under these two discrete symmetries is summarized in \tref{tab:discrsym}. 

\begin{table}
\caption{Discrete symmetries of TIM.} \label{tab:discrsym} 
\begin{indented}
  \item[]\begin{tabular}{@{}l l l}
  \br
    field & spin-reversal & Kramers--Wannier  \\[3pt] 
    \mr
    $\varepsilon$ & $\varepsilon$ & $-\varepsilon$ \\
    $t$ & $t$ & $t$ \\
    $\varepsilon''$ & $\varepsilon''$ & $-\varepsilon''$ \\
    $\sigma$ & $-\sigma$ & $\mu$ \\
    $\sigma'$ & $-\sigma'$ & $\mu'$ \\
    \br        
  \end{tabular}
\end{indented}
\end{table}

The tricritical Ising model can also be realized in terms of a coset construction of a Wess--Zumino--Witten model on the group
\begin{eqnarray}
  \frac{(\group{E}_7)_1\otimes(\group{E}_7)_1}{(\group{E}_7)_2} \,.
\end{eqnarray} 
(see \cite{IJMPA11:5327} where this property is exploited for the computation of the matrix elements of the energy-momentum tensor as well as of its two-point correlation function). 

The tricritical Ising model is also the simplest example of superconformal field theory \cite{PLB151:37}: its Hilbert space contains a finite number of irreducible representations of the super-Virasoro algebra (the antiholomorphic part is omitted)
\begin{eqnarray*}
  [L_m,L_n]&=&(m-n)L_{m+n}+\frac{c}{12}(m^3-m)\delta_{m,-n} \\
  \{G_m,G_n\}&=&2L_{m+n}+\frac{c}{3}\left(m^2-\frac{1}{4}\right)\delta_{m,-n} \\
  {[L_m,G_n]}&=&\left(\frac{1}{2}m-n\right)G_{m+n} \,. 
\end{eqnarray*}
As the generators $L_n$ arise as coefficients in the expansion of the energy-momentum tensor, so the $G_n$ are Fourier components of the superpartner of the energy-momentum tensor, the field $G(z)$ with conformal dimensions $(3/2,0)$. The theory is splitted into two sectors depending on the boundary conditions imposed on the $\Im(w)$ direction: in the Neveu--Schwarz sector the even fields can be grouped into a superfield
\begin{eqnarray}
  \mathcal{N}(z,\bar{z},\theta,\bar{\theta})=\varepsilon(z,\bar{z})+\bar{\theta}\psi(z,\bar{z})+\theta\bar{\psi}(z,\bar{z})+\theta\bar{\theta}t(z,\bar{z}) \,,
\end{eqnarray}
where $\theta,\bar{\theta}$ are Grassman variables; in the Ramond sector the magnetic fields give rise to two irreducible representations. 

\subsection{Off-critical behaviour} \label{sec:offcrit}

This section contains a brief summary of what happens when we depart from the fixed point along one of the four relevant directions. The perturbed theories are formally defined by the actions
\begin{eqnarray}
  \mathcal{A}_{i}=\mathcal{A}_{(c=7/10)}+g_i\int \rmd^2x\, \varphi_i(x) \qquad i=1,\dots,4 \,,
\end{eqnarray} 
where $g_i$ can be either positive or negative.
Since $\varphi_1$ and $\varphi_3$ are odd under spin-reversal transformation, a change of the sign of $g_1$ or $g_3$ has no effect on the spectrum: we will simply write $\mathcal{A}_1$ or $\mathcal{A}_3$. In a similar way, $\varphi_2$ is even with respect to the spin-reversal transformation, but odd under Kramers--Wannier duality, hence $\mathcal{A}_{2}^+$ and $\mathcal{A}_{2}^-$ are dual descriptions of the high- and low-temperature regimes of the same theory. Only $\varphi_4$ is even with respect to both the $\mathbb{Z}_2$ symmetries of TIM: as a consequence, $\mathcal{A}_{4}^+$ and $\mathcal{A}_{4}^-$ represent two physically distinct theories. A quick reference to the main features of these theories is provided by \tref{tab:defqfts}.

\begin{table}
 \caption[Properties of the QFTs obtained as deformations of TIM]{Properties of QFTs obtained by deformation of tricritical Ising model} \label{tab:defqfts} 
 \begin{indented}
  \item[]\begin{tabular}{@{}l l l l l l}
  \br
   $\mathcal{A}_1$ & $\mathcal{A}_{2}^+$ & $\mathcal{A}_{2}^-$ & $\mathcal{A}_3$ & $\mathcal{A}_{4}^+$ & $\mathcal{A}_{4}^-$ \\ 
   \mr
   nonintegr. & integr. & integr. & integr. & integr. & integr. \\
   3 masses & $\group{E}_7$ & $\group{E}_7$ & kinks & massless & SuSy kinks \\
   & high-temp. & low-temp. & & & \\
  \br
  \end{tabular}
 \end{indented} 
\end{table}
   
The off-critical behaviour of the tricritical Ising model is described in great detail in \cite{NPB348:591,PRE63:016103}, from which all the following data are borrowed. 

\subsubsection{Leading magnetic perturbation}

The perturbation by the field $\varphi_1\equiv\sigma$ breaks all the symmetries of the conformal theory. For this deformation, Zamolodchikov's counting argument doesn't suggest integrability, moreover a TCS analysis shows that lines repel each other in the crossover region \cite{NPB348:591}: for these reasons we are reasonably confident that the off-critical behaviour is not integrable. Because of the lack of integrability, a numerical approach like the truncation of conformal space is the only source of data about this theory.    

The spectrum contains three stable masses below threshold:
\begin{eqnarray}
  m_1 \qquad m_2=1.6(2)m_1 \qquad m_3=1.9(8)m_1 \,,
\end{eqnarray}
where the digit between round brackets is affected by error. The relation between the fundamental mass and the coupling constant $g_1$ has been estimated to be
\begin{eqnarray}
  m_1\approx 3.242\dots g_1^{40/77} \,.
\end{eqnarray}

\subsubsection{Leading energy perturbation}

The perturbation by $\varphi_2\equiv\varepsilon$ with positive coupling constant drives the system into its high-temperature phase. The off-critical theory is integrable and related to the Toda field theory based on the exceptional algebra $\group{E}_7$: the conserved currents have spins
\begin{eqnarray}
  s=1,5,7,9,11,13,17 \pmod{18} \,,
\end{eqnarray}  
(these numbers are the Coxeter exponents of the exceptional algebra $\group{E}_7$) and the spectrum is given in \tref{tab:e7}.  

\begin{table} 
  \caption{Spectrum of the theory $\mathcal{A}_2^+$.} \label{tab:e7} 
  \begin{indented}
  \item[]\begin{tabular}{@{}l l l}
  \br
  exact & numerical & parity \\
  \mr
   $m_1$ & 1 & odd \\
   $m_2=2m_1\cos(5\pi/18)$ & 1.2856\dots & even \\
   $m_3=2m_1\cos(\pi/9)$ & 1.8794\dots & odd \\
   $m_4=2m_1\cos(\pi/18)$ & 1.9696\dots & even \\
   $m_5=4m_1\cos(\pi/18)\cos(5\pi/18)$ & 2.5321\dots & even \\
   $m_6=4m_1\cos(2\pi/9)\cos(\pi/9)$ & 2.8794\dots & odd \\
   $m_7=4m_1\cos(\pi/18)\cos(\pi/9)$ & 3.7017\dots & even \\
  \br
  \end{tabular}
 \end{indented}
\end{table}

The $S$-matrix, as well as the particle masses, is known exactly \cite{NPB330:465,IJMPA5:1025}. The relationship between the mass gap and the coupling constant can as well be exactly computed \cite{PLB324:45}:
\begin{eqnarray}
  m_1&=&\left(\frac{2\Gamma(\frac{2}{9})}{\Gamma(\frac{2}{3})\Gamma(\frac{5}{9})}\right)
  \left(\frac{4\pi^2\Gamma(\frac{2}{5})\Gamma^3(\frac{4}{5})}{\Gamma^3(\frac{1}{5})\Gamma(\frac{3}{5})}\right)^{5/18} |g_2|^{5/9} \\ 
  &=&|g_2|^{5/9} 3.7453728362\dots \,. \nonumber
\end{eqnarray}

Since this theory is integrable, one can exploit the form factor expansion in order to compute correlation functions. However, as observed in \cite{PRE63:016103}, an obstacle occurs that drastically reduces the precision one can achieve by this technique: the asymptotic factorization argument \cite{NPB455:724} fails to discriminate between the form factors of $\varphi_1$ and those of $\varphi_3$. The one-particle form factors have therefore to be determined by some independent way, like the truncation of conformal space introduced in \sref{sec:TCS}. 

\subsubsection{Subleading magnetic perturbation}

Also the perturbation by the field $\varphi_3$ is integrable \cite{IJMPA6:1407}. The presence of two degenerate (and asymmetrical) vacua permits the existence of two massive kink excitations and one breather bound state, all with the same mass. The $S$-matrix is exactly known \cite{IJMPA7:5281}, as well as the relationship between the mass gap and the coupling constant:
\begin{eqnarray}
  m_1 &=& \frac{\sqrt{3}\Gamma(\frac{1}{3})\Gamma(\frac{5}{9})}{\pi\Gamma(\frac{8}{9})}
  \left[\frac{\pi^2\Gamma(\frac{1}{4})\Gamma^2(\frac{11}{16})}
  {\Gamma(\frac{3}{4})\Gamma^2(\frac{5}{16})}\right]^{4/9}g_3^{8/9} \\
   &=& g_3^{8/9} 4.92779064\dots \,. \nonumber
\end{eqnarray}

\subsubsection{Subleading energy perturbation}

The deformation by the field $\varphi_4$ is related to the change in the vacancy density: if we move in the direction $g_4>0$, then we are increasing the number of spins that take values $\pm1$, while in the direction $g_4<0$ the spins taking value 0 are favored. Both the theories are integrable \cite{NPB358:497}. The $g_4>0$ deformation originates a massless renormalization group flow \cite{NPB316:590} at the ending point of which there is the Ising field theory $\mathcal{M}(4,3)$. Along this flow, the conformal dimension of the magnetization operator changes from the value $\frac{3}{80}$ to $\frac{1}{16}$, while the conformal dimension of the energy operator changes from $\frac{1}{10}$ to $\frac{1}{2}$. The exact massless $S$-matrix \cite{NPB358:524} and form factors \cite{PRD51:6620} for the theory $\mathcal{A}_4^+$ are available. If $g_4<0$, on the other hand, the corresponding Landau--Ginzburg potential is threefold degenerate: the elementary excitations are massive kinks \cite{NPB358:497}. The scattering theory is considered in \cite{NPB554:537}.  

\section{Free energy of TIM} \label{sec:FE}

This section collects the results of our research about the free energy of the tricritical Ising model. After introducing our notations in \sref{sec:conv}, we discuss the analytic structure of the theory simultaneously deformed by the leading magnetic and thermic perturbations (the case more similar to the Ising field theory) in \sref{sec:real} and \sref{sec:compl}. The effect of turning on the sub-leading energy perturbation is treated in \sref{sec:3slice}. Some interesting byproducts of our numerical data are collected in the last section.  

\subsection{Conventions} \label{sec:conv}

In this paper we are concerned with the quantum field theory formally defined on a cylinder of circumference $R$ with periodic boundary conditions by the action
\begin{equation} \label{eq:action}
  \mathcal{A}[g_1,g_2,g_3,g_4]=\mathcal{A}_{(c=7/10)}
  +\sum_{i=1}^4 g_i\int_{-\infty}^\infty du \int_0^R dv\,\varphi_i(u,v) \,,
\end{equation}
where the fields $\varphi_i(u,v)$ are defined in \tref{tab:TIM}. We will outline location and nature of the singularities of the free energy by considering two- or three-dimensional slices of the total four-dimensional parameter space. When considering such a slice, we will write explicitly only those couplings that are different by 0: for instance, 
\begin{equation}
  \mathcal{A}[g_1,g_2,0,0]\equiv\mathcal{A}[g_1,g_2] \,.
\end{equation}
Sometimes it is important to stress that, say, $g_2$ is negative or $g_1$ is imaginary. In such occasions, $|g_i|$ is denoted by the appropriate one among the following symbols:
\begin{equation}
  h\equiv|g_1| \qquad \tau\equiv|g_2| \qquad h'\equiv|g_3| \qquad \chi\equiv|g_4| \,.
\end{equation}  
To give an example, $\mathcal{A}[-\tau,ih']$ means $\mathcal{A}[g_2,g_3]$ with $g_2$ real and negative, $g_3$ purely imaginary with positive imaginary part. 

The same convention applies to the correspondent Hamiltonian:
\begin{equation} \label{eq:ham}
  \hat{H}[g_1,g_2,g_3,g_4]=\hat{H}_0+\sum_{i=1}^4 g_i\int_0^Rdv\,\hat{\varphi}_i(u,v) \,, 
\end{equation}
where the hat indicates quantities defined on the cylinder:
\begin{eqnarray} 
  \hat{H}_0&=&\frac{2\pi}{R}\left(L_0+\bar{L}_0-\frac{c}{12}\right) \\
  \hat{\varphi}_i&=&\left|\frac{2\pi}{R}\right|^{2h_i}\varphi_i \,.
\end{eqnarray}

Several dimensionless ratios between the coupling constants are used in this chapter: it is useful to collect all the definitions here for a quick reference:
\begin{eqnarray}
  \xi=\frac{g_1}{|g_2|^{77/72}} & \qquad &   
  \eta'=\frac{g_2}{|g_1|^{72/77}}   \label{eq:defxi}  \\
  \zeta'=\frac{g_4}{|g_1|^{32/77}} & \qquad & 
  \eta=\frac{g_2}{g_1^{72/77}}  \label{eq:defzeta} 
\end{eqnarray}
Throughout this section, we assume $g_2,g_4\in\mathbb{R}$: it is sensible to study the analytic properties of the free energy in complex temperature~\cite{LTP7C:1,JMP7:2000} or chemical potential, but we don't want to treat this problem now. The magnetic field $g_1$ is assumed to be real in the next section, and complex in \sref{sec:compl}. It would have been equally interesting to study the analyticity in the subleading magnetic field $g_3$, but since $h_3=7/16$, the ultraviolet divergences quickly spoils the TCS of any attendibility, at least as far as the evaluation of the ground state is concerned.  

\subsection{Free energy of $\mathcal{A}[g_1,g_2]$: real couplings} \label{sec:real} 

We will in this section begin the systematic exploration of the analytic properties of the free energy. The Hamiltonian $H[g_1,g_2]$ appears to be the best starting point for at least two reasons: first, the TCS produces the most accurate results when the most relevant perturbations are present; moreover, the affinity with the Ising field theory can provide a valuable source of inspiration. Let us then consider the doubly deformed Hamiltonian
\begin{equation} \label{eq:Hg1g2}
  H[g_1,g_2]=\frac{2\pi}{R}\left[H_0+2\pi g_1\left(\frac{R}{2\pi}\right)^{77/40}V_1
  +2\pi g_2\left(\frac{R}{2\pi}\right)^{9/5}V_2\right] \,.
\end{equation}
We assume $g_2\neq 0$ and fix $|g_2|^{5/9}$ as the mass unit of measure. Then we introduce dimensionless Hamiltonian $\mathcal{H}_\pm=H/|g_2|^{5/9}$ and cylinder radius $r=R|g_2|^{5/9}/(2\pi)$, so that we can write
\begin{equation} \label{eq:H12}
  \mathcal{H}_{\pm}(\xi)=\frac{1}{r}\left[H_0+2\pi\xi r^{77/40} V_1\pm2\pi r^{9/5}V_2\right] \,
\end{equation}
where the sign is plus in the high-temperature regime $g_2>0$ and minus in the low-temperature one $g_2<0$, while $\xi$ is defined by \eref{eq:defxi}, with $g_1$ real.
 
The free energy density, that can be extracted from the behaviour of the ground state of the Hamiltonian~\eref{eq:Hg1g2} by taking the infrared limit $R\to\infty$, has dimensionality of a squared mass, therefore can be written
\begin{equation} \label{eq:defG}
  F(g_1,g_2)=|g_2|^{10/9}(\mathcal{E}_0+\mathcal{F}(\xi)) \,,
\end{equation} 
where $\mathcal{E}_0$ is exactly known~\cite{PLB324:45}
\begin{eqnarray}
   \mathcal{E}_0&=&-\frac{\sin(2\pi/9)}{8\sin(\pi/3)\sin(5\pi/9)}\mathcal{C}_2^2
   =-1.32155588\dots \\
   \mathcal{C}_2&=&\frac{2\Gamma(2/9)}{\Gamma(2/3)\Gamma(5/9)}\left( 4\pi^2 \frac{\Gamma(2/5)}
   {\Gamma(3/5)} \left( \frac{\Gamma(4/5)}{\Gamma(1/5)} \right)^3 \right)^{5/18} \,,
\end{eqnarray}
and $\mathcal{F}(\xi)$ is the dimensionless free energy density. 

For practical and conceptual reasons, it is convenient to split the space of couplings into high- and low-temperature sectors and treat them separately. In the region $g_2>0$, indeed, the keywords are distribution of zeros of partition function and Yang--Lee (in fact, Yang--Lee-like) edge singularity, while for $g_2<0$ the relevant physical ideas are metastable vacuum decay and spinodal singularity. These two regimes, however, are deeply related, and we will show how one can make full use of this relationship. 

In the high-temperature regime, $\mathcal{F}_\mathrm{high}(\xi)$ is even: $\mathcal{F}_\mathrm{high}(\xi)=\mathcal{F}_\mathrm{high}(-\xi)$, as a consequence of the $\mathbb{Z}_2$ spin-reversal symmetry. No phase transition is present for $g_2>0$, hence we can expand $\mathcal{F}$ around $\xi=0$ writing the convergent series 
\begin{equation} \label{eq:Fhigh}
  \mathcal{F}_\mathrm{high}(\xi)=\mathcal{F}_2\xi^2+\mathcal{F}_4\xi^4+\mathcal{F}_6\xi^6+\dots \,.
\end{equation}
The coefficient $\mathcal{F}_2$ is proportional to the high-temperature susceptibility at $h=0$ that can be estimated either by integrating the two-spin correlation function
\begin{equation}
   \int d^2x <\varphi_1(x)\varphi_1(0)>_c \,,
\end{equation}
or by using the TCS. Our best estimate, obtained by fitting the TCS spectra for different values of $\xi$, is 
\begin{equation}
   \mathcal{F}_2=-\frac{\Gamma_{11}^{2+}}{2}=-0.046(9) \,,
\end{equation}
where the notation $\Gamma_{11}^{2+}$ for the susceptibility is borrowed from \cite{PRE63:016103}. As far as we know, no data about higher order coefficients is available at the moment. In principle, one could use form factor expansion of the four-spin correlation function in order to determine $\mathcal{F}_4$, like the authors of \cite{NPB583:614} did for the Ising model, but no attempt has been done of computing three-particle form factors for the tricritical Ising model\footnote{The main obstacle is that the error associated to the one- and two-particle form factors propagate while solving the system of linear equations needed to build the three-particle ones. While for the Ising case one can determine the lowest form factors with the desired precision, this is not possible for TIM~\cite{PRE63:016103}.}. 

In the low-temperature regime, if one tries to smoothly vary the magnetic field from positive values to  negative ones, at $\xi=0$ one has residual magnetization which does not disappear immediately as $\xi$ becomes negative. The system enters a metastable phase in $\xi=0$, where a weak singularity is predicted by Langer theory. The spin-reversal symmetry being spontaneously broken, the function $\mathcal{F}_\mathrm{low}(\xi)$ is not even: its asymptotic expansion reads
\begin{equation}
  \mathcal{F}_\mathrm{low}(\xi)=\widetilde{\mathcal{F}}_1\xi+\widetilde{\mathcal{F}}_2\xi^2+\widetilde{\mathcal{F}}_3\xi^3+\dots \,.
\end{equation}
The first coefficient $\widetilde{\mathcal{F}}_1$ is the spontaneous magnetization at $g_1=0$: it can be exactly evaluated~\cite{NPB516:652}
\begin{equation}
   \widetilde{\mathcal{F}}_1=-B_{12}=-1.59427\dots \,.
\end{equation}
Again, the conventions of \cite{PRE63:016103} are assumed.  
The second coefficient $\widetilde{\mathcal{F}}_2$ is related to the low-temperature susceptibility $\Gamma_{11}^{2-}$. Our estimate (see \tref{tab:compamp} for a comparison with results obtained by means of other approaches) is 
\begin{equation}
   \widetilde{\mathcal{F}}_2=-\frac{\Gamma_{11}^{2-}}{2}=-0.011(8) \,.
\end{equation}
No other term is available at the moment. Also in this case, the determination of further coefficients requires the computation of multi-spin correlation functions. 

\subsection{Free energy of $\mathcal{A}[g_1,g_2]$: complex magnetic field} \label{sec:compl}

In this section we analytically continue the magnetic field $g_1$ (and therefore $\xi$) to complex values. The Hamiltonian to be diagonalized is still given by \eref{eq:H12}, but its eigenvalues are in general complex numbers. For the Yang--Lee theorem, we expect the free energy $\mathcal{F}(\xi)$ to be analytic in some neighborhood of $\xi=0$ in the high-temperature plane; in the low-temperature plane, if the analytic continuation starts from the positive real axis, we expect analyticity at least in the right half-plane. It is useful to keep in mind, for a comparison, the results of Fonseca--Zamolodchikov's analysis of the Ising model, summarized in \fref{fig:HTIsing} and \fref{fig:LTIsing}. 

\subsubsection{High-temperature regime} \label{sec:HT} 

The first problem to solve is to determine the convergence radius of the expansion \eref{eq:Fhigh}, that amounts to locate the singularities created by the accumulation of zeros of the partition function. The distribution of Yang--Lee zeros in the Blume--Capel model has been studied for the first time by Suzuki \cite{PTP40:1246} (see also \cite{PRB7:3141,PRB9:2030}). He proved that all the logarithmic singularities of the free energy are located on the imaginary axis of the complex $H$-plane, just like in the Ising model, provided that $\beta\Delta<\ln 2$, that is, loosely speaking, if the dilute Ising model is not too dilute. For $\beta\Delta>\ln 2$, the zeros lie on arcs that can possibly have no intersection at all with the imaginary axis \cite{PRL84:4794,JSP116:97}. In the mean-field approximation, the tricritical point is characterized by $\beta_t\Delta_t=2\ln 2$, hence should be out of the region where Suzuki's result is valid. However, the mean-field approximation should not be taken too seriously in two dimensions, so we have better to guess the position of the singularities of the free energy by carefully inspecting the TCS results without prejudices.

If we look back at the Hamiltonian \eref{eq:Hg1g2}, we easily realize that $\xi$ is not the only dimensionless ratio between couplings we can define. Indeed, we can decide to measure the masses in units of $|g_1|^{40/77}$, so that instead of Eq.~(\ref{eq:H12}) we get
\begin{equation} \label{eq:H'}
  \mathcal{H}'_{\arg(g_1)}(\eta')=\frac{1}{r'}\left[H_0+2\pi e^{i\arg(g_1)}{r'}^{77/40}V_1
  +2\pi \eta' {r'}^{9/5} V_2\right] \,,
\end{equation}
where 
\begin{equation} \label{eq:defeta'}
  \eta'=\frac{g_2}{|g_1|^{72/77}} \,.
\end{equation}
Note that with this definition $\eta'$ is always real, even for complex $g_1$. Moreover, in this variable the high- and low-temperature regimes are analyzed together. The spectrum obtained by diagonalizing the Hamiltonian \eref{eq:H'} is obviously related to the one extracted from \eref{eq:H12} by means of the relations
\begin{equation}
  E'=E{\eta'}^{5/9}  \qquad \qquad r=r'{\eta'}^{5/9} \,.
\end{equation} 
The spectrum of the Hamiltonian $H[g_1,g_2]$ can thus be studied in terms of $\xi$ in the region where $|g_1|$ is small, while $\eta'$ provides access to the spectrum for large values of $|g_1|$: by combining the two approaches we have been able to explore the entire range of $g_1$.

\begin{figure}
 \centering
 \mbox{\subfigure[$\eta'=0.06$: ground state real]{\epsfig{figure=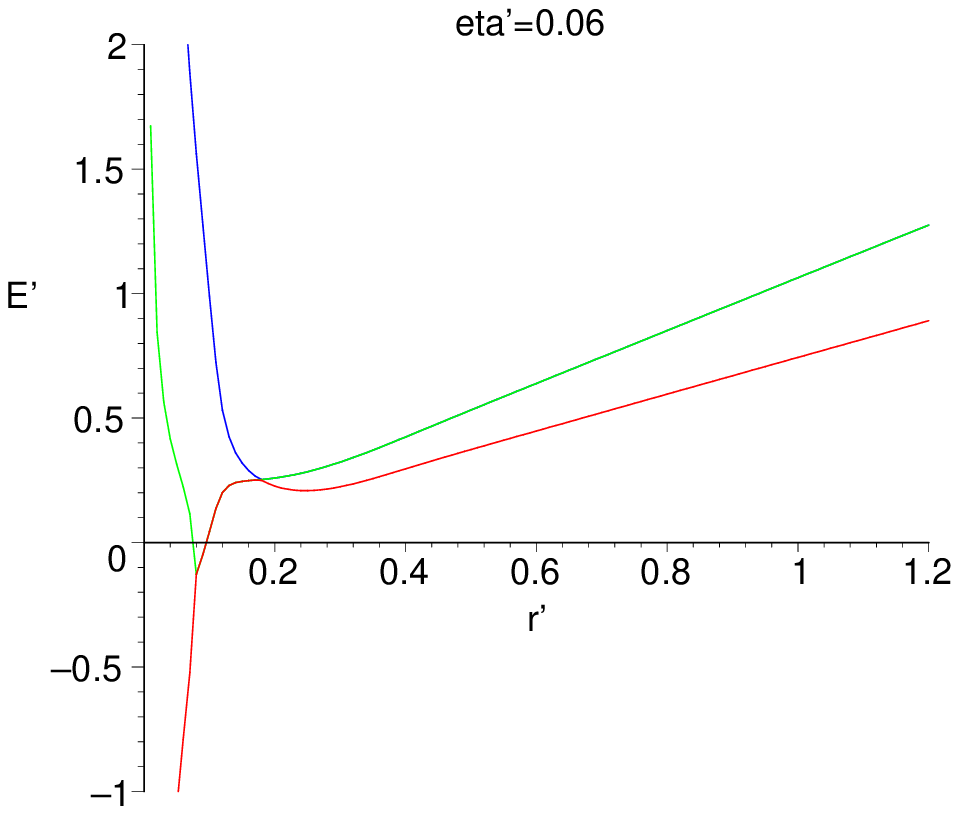, width=6cm}}\quad
       \subfigure[$\eta'=0.05$: ground state real]{\epsfig{figure=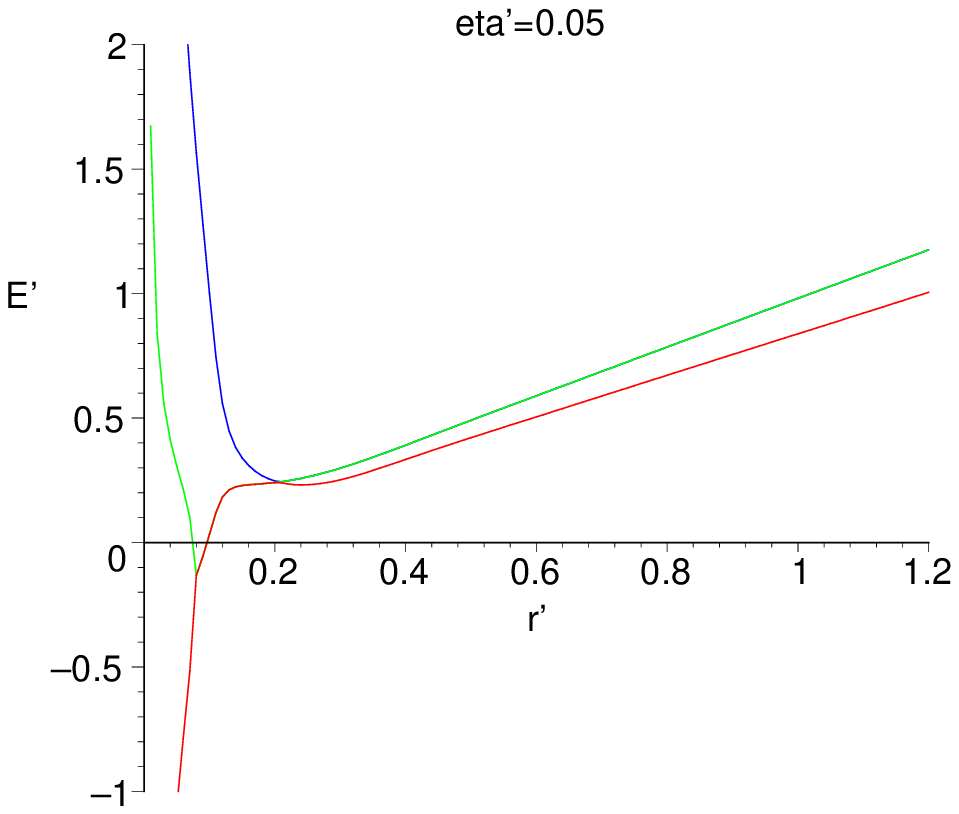, width=6cm}}}
 \mbox{   
       \subfigure[$\eta'=0.04$: ground state complex]{\epsfig{figure=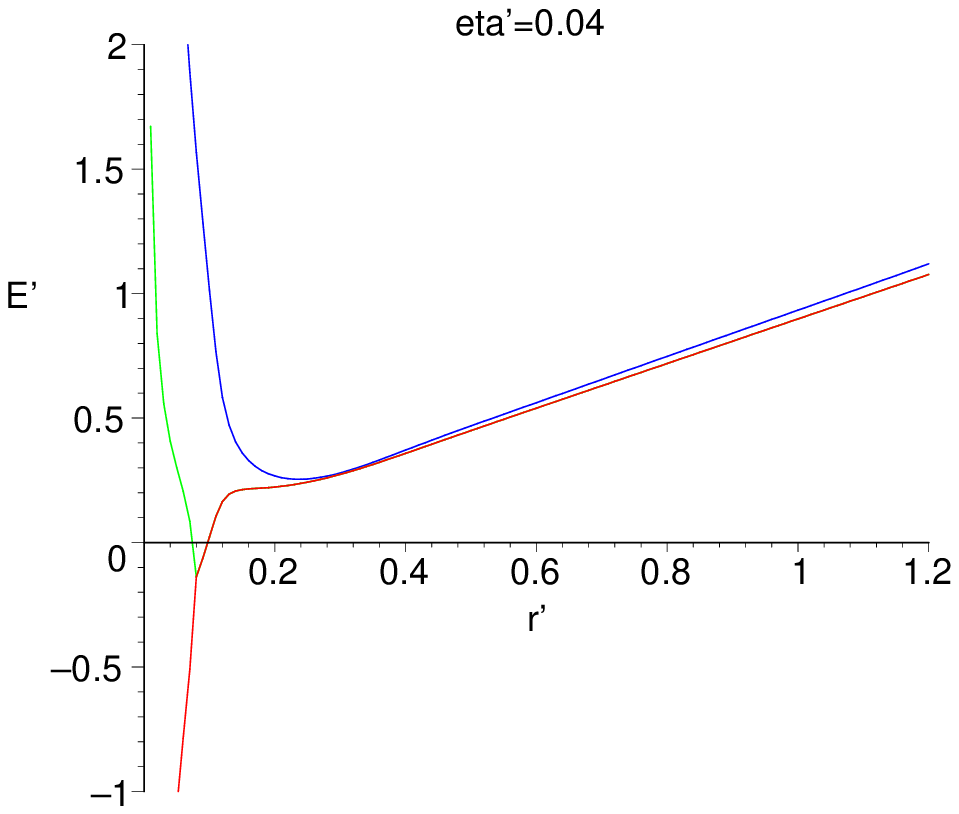,width=6cm}}
       \subfigure[$\eta'=0.03$: ground state complex]{\epsfig{figure=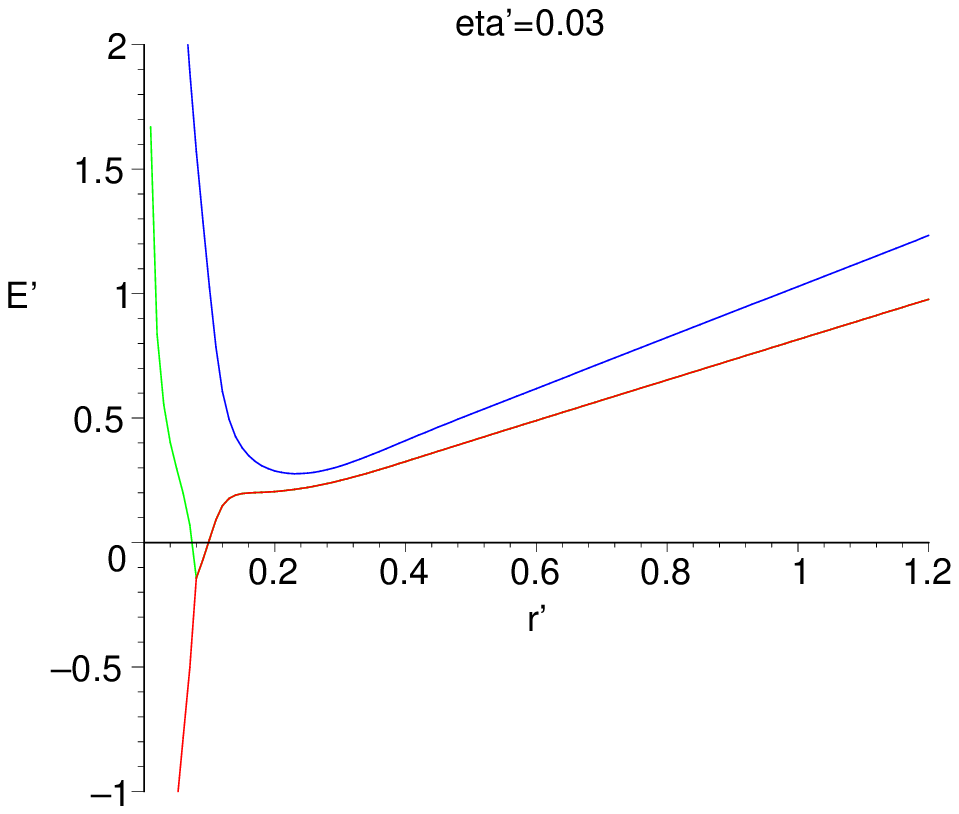,width=6cm}}}
  \caption[Spectrum of $\mathcal{H}'_{\pi/2}(\eta')$]{Spectrum of $\mathcal{H}'_{\pi/2}(\eta')$: real parts of the three lowest eigenvalues. When two lines overlap because they are forming a conjugate pair, only the color of the lowest one is showed. The critical value is $0.0419<\eta'_\mathrm{c}<0.0420$.} \label{fig:spec12c}
\end{figure}

Inspired by the Ising case, we looked for the presence of singularities on the imaginary axis of the $\xi$ plane ($\arg(g_1)=\pi/2$). \Fref{fig:spec12c} shows the real part of the three lowest levels of the spectrum of $\mathcal{H}'_{\pi/2}$ for different values of $\eta'$. Few explanations are in order here. The analytic continuation of $\xi$ to complex values does not spoil the spin-reversal symmetry, hence the spectrum is still invariant under the transformation $\xi\to-\xi$. When $\xi$ is purely imaginary, we have $-\xi=\xi^*$ so the $\mathbb{Z}_2$ symmetry becomes invariance of the spectrum under complex conjugation. Therefore the eigenvalues of $\mathcal{H}'_{\pi/2}$ must be real, or come in complex conjugate pairs. It is easy to distinguish the region $\eta'>\eta'_\mathrm{c}$, where (for values of $R$ within the physical window) the ground state is real while the second and third levels form a complex conjugate pair, from the region $\eta'<\eta'_\mathrm{c}$, where the ground state is complex and the third level is real. The critical value is $0.0419<\eta'_\mathrm{c}<0.0420$. One can check that $\mathcal{H}_+(\xi)$ shows the same transition for 
\begin{equation} \label{eq:defxic}
29.67\rmi<\xi_\mathrm{c}<29.74\rmi \,. 
\end{equation} 
A careful inspection of the spectra obtained for values of $\xi$ in the right half-plane shows no evidence of other singular point (except, of course, the complex conjugate of $\xi_\mathrm{c}$). This confirms our expectations based upon the affinity between $\mathcal{A}[g_1,g_2]$ and $\mathcal{A}_\mathrm{IFT}$, defined in \eref{eq:Aift}. We can therefore cut the high-temperature $\xi$ plane just in the same fashion as the Ising model (see \fref{fig:tagli1}(b)).   

Which non-unitary minimal model is related to the edge singularity that arises in TIM? To the best of our knowledge, the only published paper about this interesting question was authored by von Gehlen \cite{IJMPB8:3507}, who applied finite-size-scaling methods to a quantum chain which can be defined starting from a strongly anisotropic Blume--Emery--Griffiths model \cite{PRB23:6099}. At the end of his analysis, von Gehlen claimed that his data ``strongly hint'' that the relevant conformal field theory is the minimal model $\mathcal{M}(7,2)$. In principle, TCS gives direct access to the effective central charge through the relation, valid at the critical point, 
\begin{equation} \label{eq:ceff}
  E_0(R)=F_0R-\frac{\pi \tilde{c}}{6R}+\mathcal{O}\left(\frac{1}{R^2}\right) 
\end{equation}  
(more terms can be found in \cite{JSP110:527}). In point of fact, a great precision is needed in order to properly evaluate the subleading behavior of the ground state energy. At level 5 of truncation, it is not even possible to discriminate between $\tilde{c}=17/20$ as in $\mathcal{M}(8,5)$ and the value $\tilde{c}=4/7$ characteristic of $\mathcal{M}(7,2)$, let alone the small difference separating the latter and $\tilde{c}=3/5$ related to the third candidate $\mathcal{M}(5,3)$. 

\subsubsection{Low-temperature regime}

The functions $\mathcal{F}_\mathrm{high}(\xi)$ and $\mathcal{F}_\mathrm{low}(\xi)$ are obviously related one to another. We already introduced with \eref{eq:defeta'} a variable which describes both the high- and low-temperature sectors. However, $\eta'$ is a real variable: in order to connect the analytic structure of the high- and low-temperature $\xi$ planes we need a variable which is analytically related to $\xi$. It is sufficient to change the definition of $\eta'$ by removing the modulus in the denominator:
\begin{equation}
  \eta=\frac{g_2}{g_1^{72/77}} \,.
\end{equation}
Now we have a complex variable suitable for describing both the temperature regimes, and analytically related to $\xi$: 
\begin{equation} \label{eq:etaxihighT}
  \eta=\pm\frac{1}{\xi^{72/77}} \,,
\end{equation}
where the sign is + for the high-temperature sector and $-$ for the low-temperature one.

\begin{figure}
 \centering
 \mbox{\subfigure[$\xi=0.5$]{\epsfig{figure=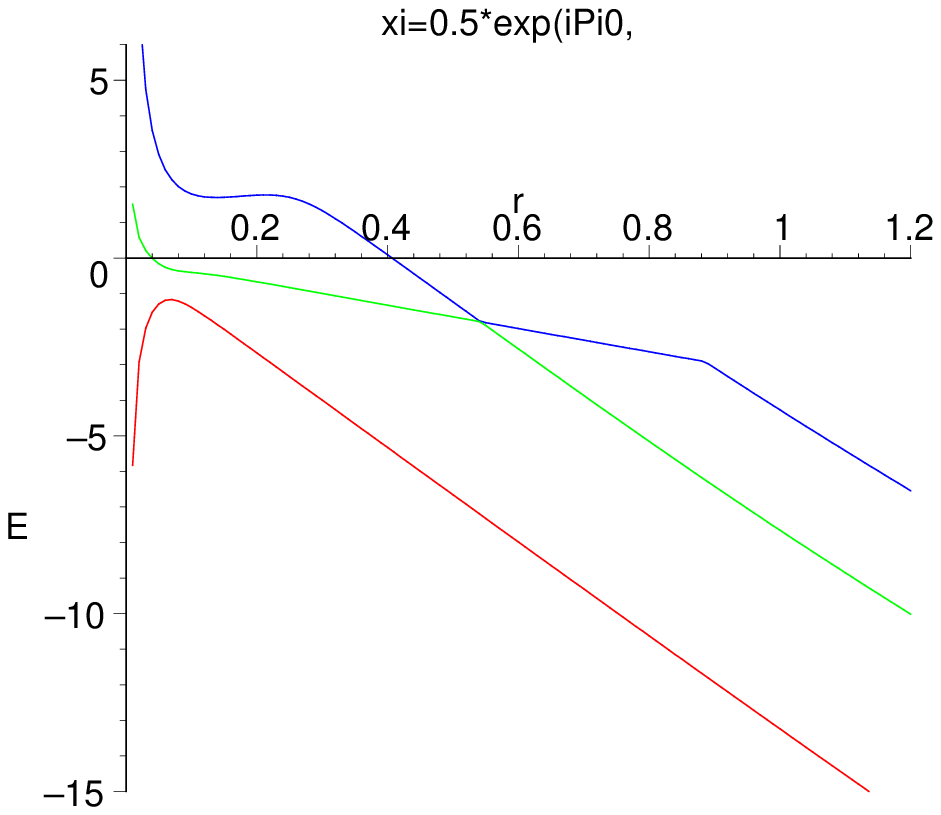, width=6cm}}\quad
       \subfigure[$\xi=0.5e^{i\pi/4}$]{\epsfig{figure=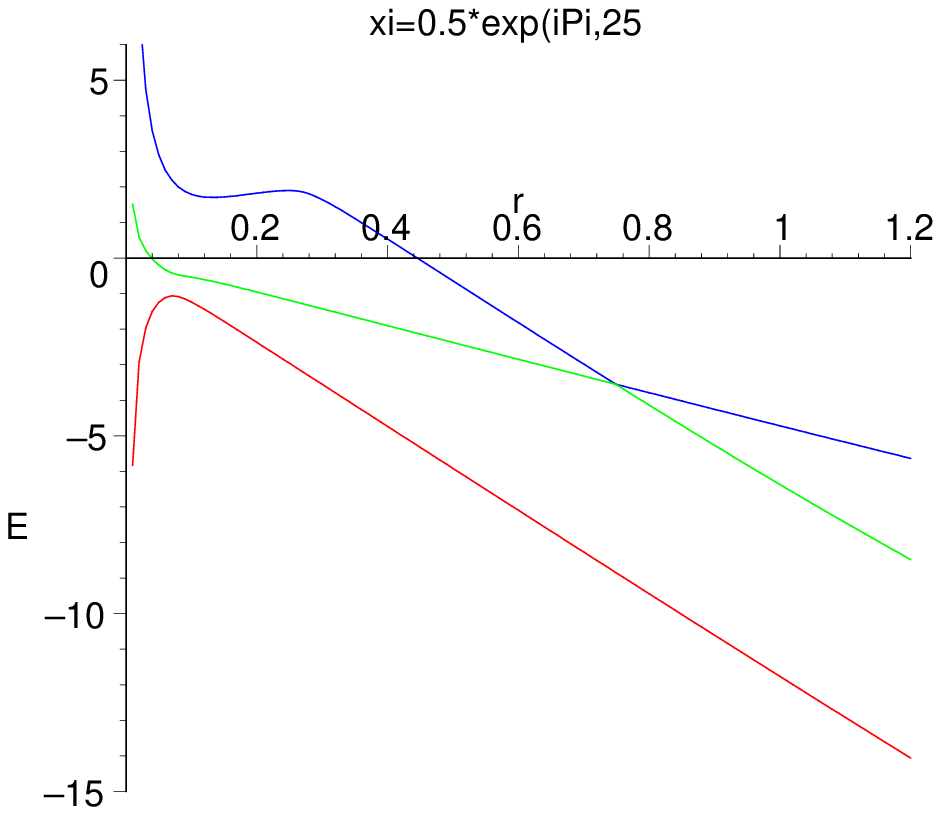, width=6cm}}}
 \mbox{   
       \subfigure[$\xi=0.5e^{i\pi9/20}$]{\epsfig{figure=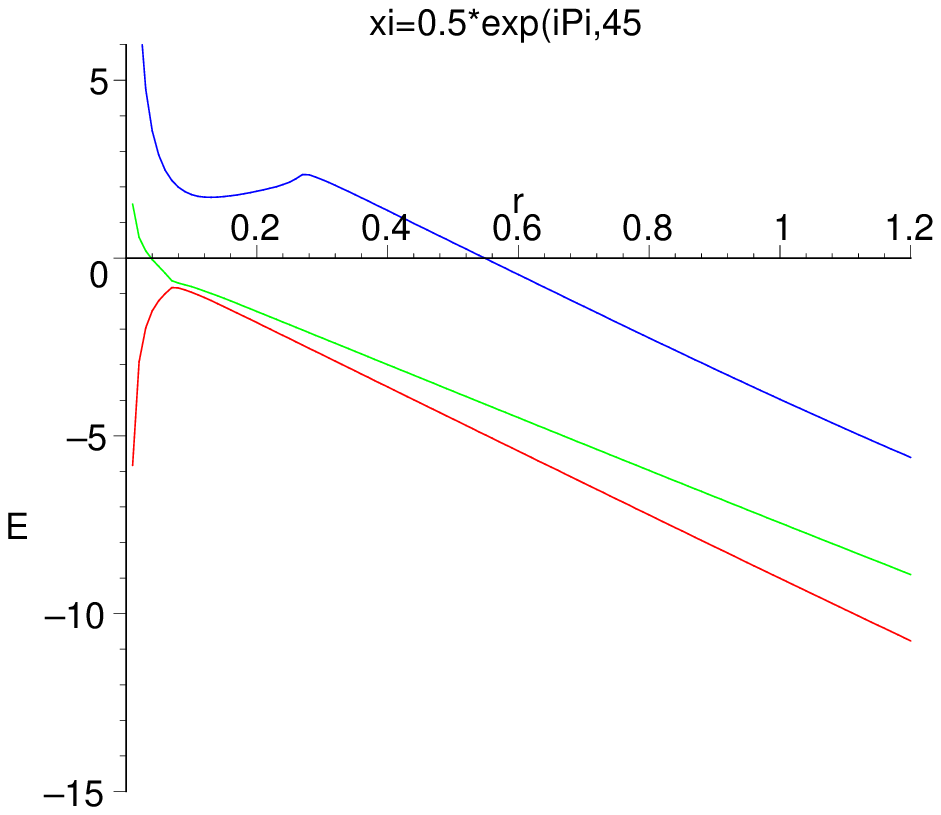,width=6cm}}
       \subfigure[$\xi=0.5e^{i\pi/2}$]{\epsfig{figure=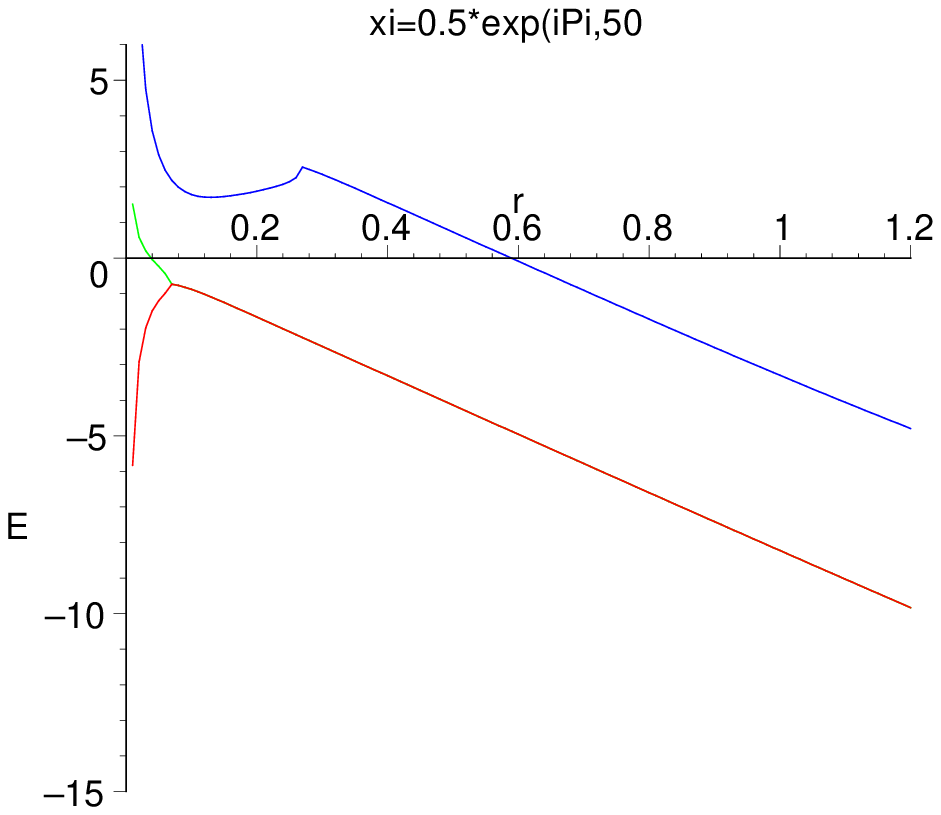,width=6cm}}}
  \caption[Spectrum of $\mathcal{H}_-(\xi)$]{Spectrum of $\mathcal{H}_-(\xi)$: real parts of the three lowest eigenvalues. As $\xi$ is rotated into the complex plane, the ground state and the false vacuum line get close, until they eventually form a complex conjugate pair when $\xi$ reaches the imaginary axis.} \label{fig:spec-21c}
\end{figure}

\begin{figure}
 \centering
 \mbox{\subfigure[Low-temperature $\xi$-plane]{\epsfig{figure=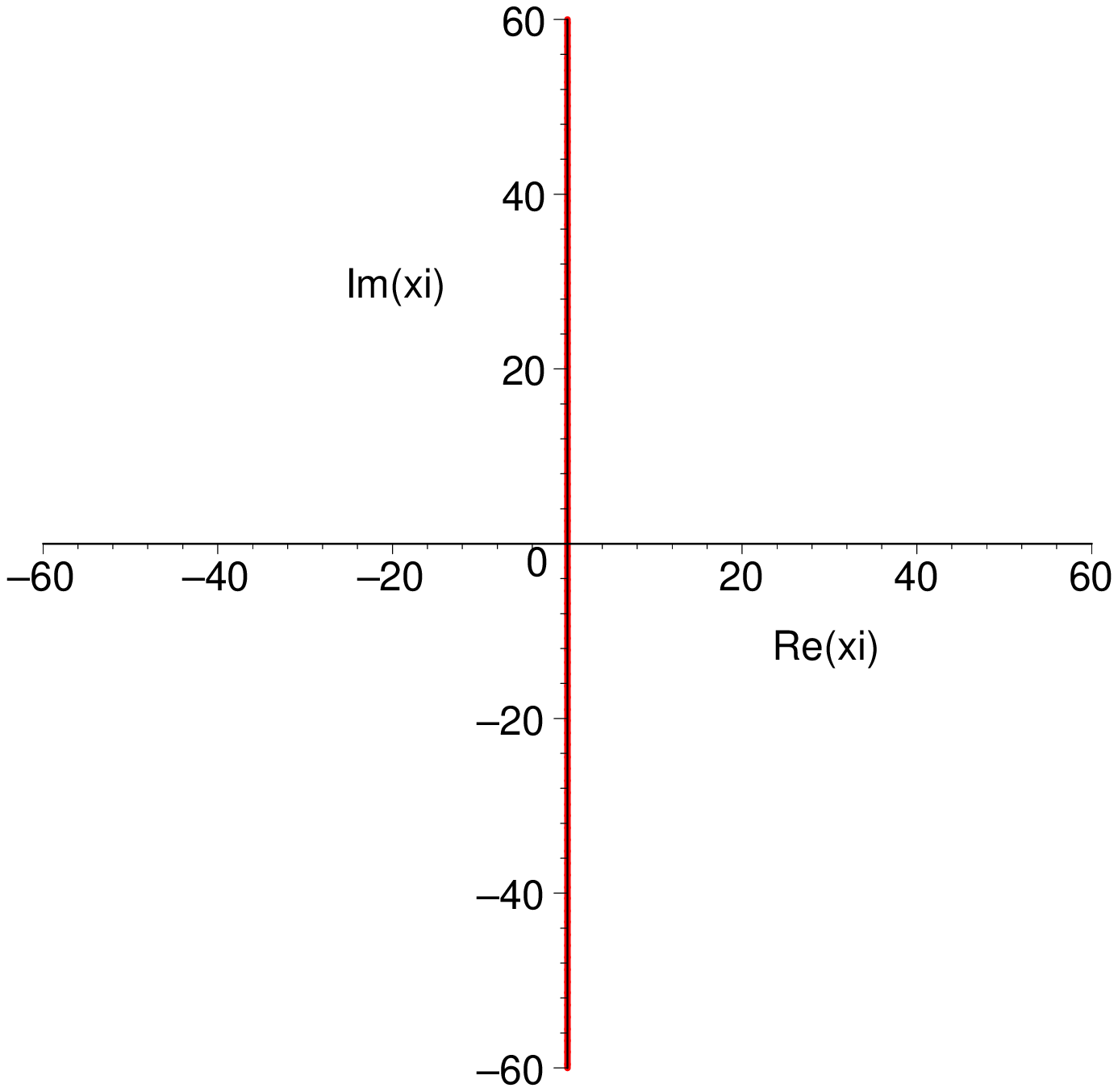, width=6cm}}
       \subfigure[High-temperature $\xi$-plane]{\epsfig{figure=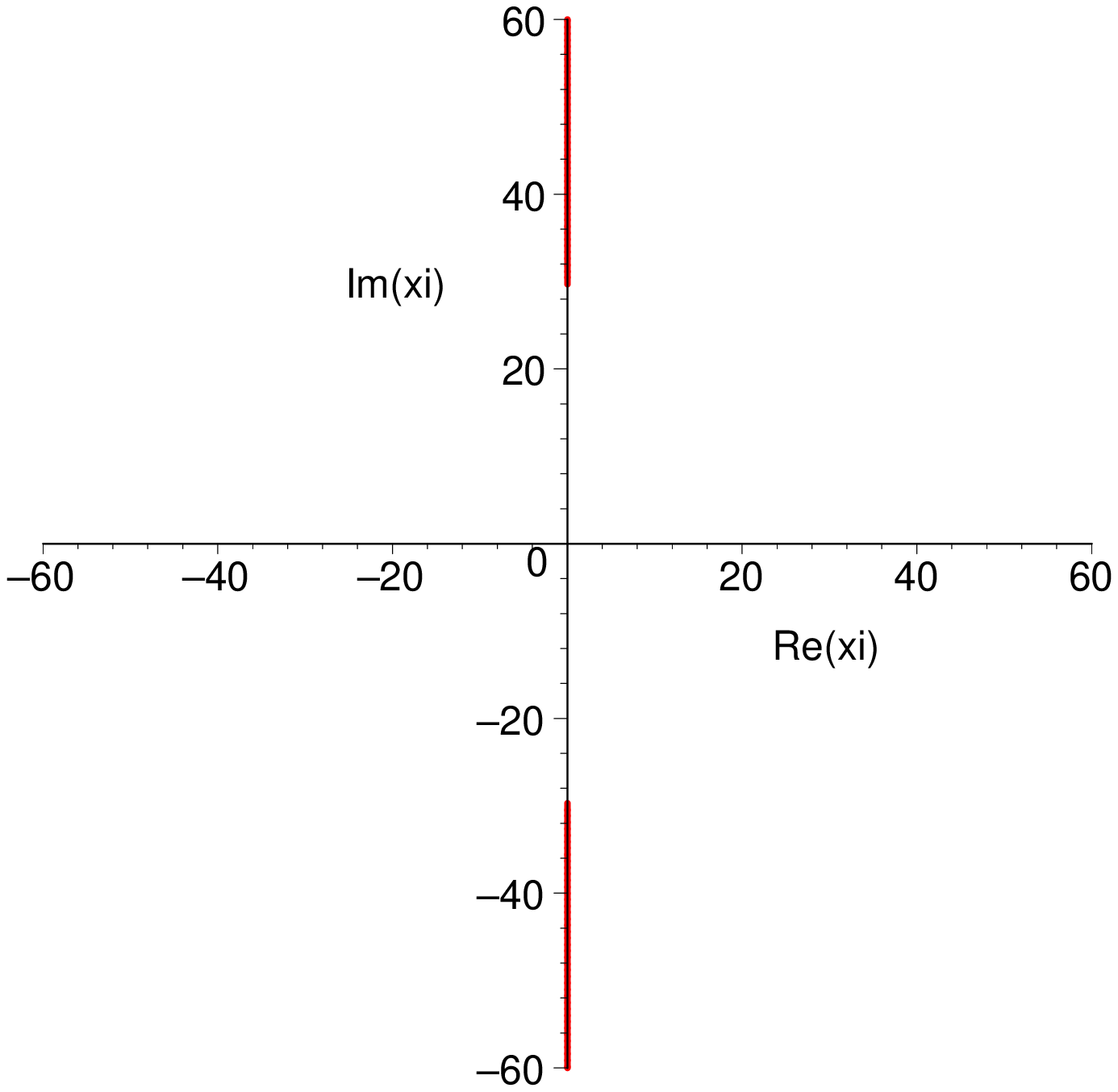, width=6cm}}}
 \mbox{\subfigure[$\eta$-plane]{\epsfig{figure=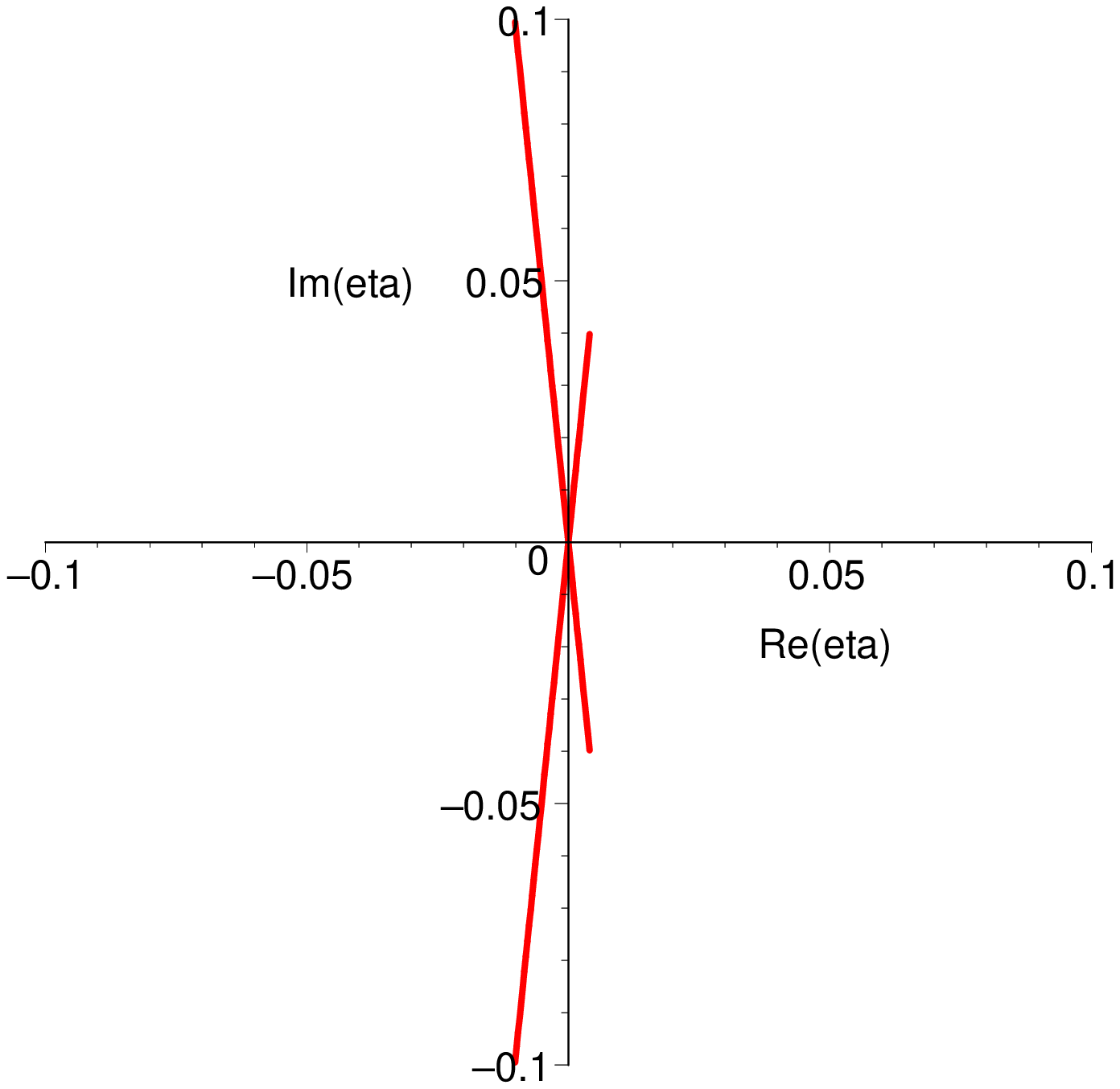,width=6cm}}}
  \caption[Cuts for the theory $\mathcal{A}(g_1,g_2)$]{The branch cuts of the theory $\mathcal{A}[g_1,g_2]$. The high-temperature right half-plane is mapped into the corner $-\frac{36}{77}\pi<\arg(\eta)<\frac{36}{77}\pi$, while the low-temperature right half-plane is mapped into the region $-\frac{36}{77}\pi<\arg(-\eta)<\frac{36}{77}\pi$.} \label{fig:tagli1}
\end{figure}

The $\eta$ plane is graphically represented in \fref{fig:tagli1}(c).
The right half-plane of the high-temperature $\xi$-plane is mapped by \eref{eq:etaxihighT} into the corner $-\frac{36}{77}\pi<\arg(\eta)<\frac{36}{77}\pi$, while the image of the right half-plane of the low-temperature $\xi$-plane is the region $-\frac{36}{77}\pi<\arg(-\eta)<\frac{36}{77}\pi$. The Yang--Lee-like branching points are mapped into the points $\eta=Y_\mathrm{c}\exp(\pm i\frac{36}{77}\pi)$, where $Y_\mathrm{c}=1/|\xi_\mathrm{c}|^{72/77}$. From \fref{fig:tagli1}(c) it is evident what we can expect when, in the low-temperature $\xi$-plane, we move from the real positive axis: we have clear way (that is, the free energy can be safely analytically continued) until we reach the imaginary axis, where we meet a branch cut that is just what in the high-temperature $\xi$-plane appears as the Yang--Lee-like branch cut (see \fref{fig:spec-21c}).


Indeed, the spectra obtained by means of TCS show that the exponential splitting amongst the first two levels, that is the effect of the non-zero amplitude of tunnelling between the two degenerate vacua, is removed as soon as an imaginary magnetic field, however small, is turned on. So the ground state is complex for any imaginary $\xi$. The TCS permits also to determine the nature of the branch cut: it is sufficient to examine the imaginary part of the ground state to realize that the values of the free energy on opposite edges of the cut are complex conjugate. This is exactly the behavior we expect on the ground of the critical droplets calculations of the metastable vacuum decay~\cite{JPA13:1755,PRB23:2317}.

The position of the singularities has physical meaning, but cutting the plane is a matter of conventions; in particular, we can, following Fonseca and Zamolodchikov, draw a cut passing through zero in the high-temperature $\xi$ plane, so that in the $\eta$ plane we have cleared the way for going into the low temperature sector while remaining close to the real axis, where $\mathcal{F}(\eta)$ is analytic. On the low-temperature $\xi$ plane, this operation removes the branch cut along the imaginary axis, and we can extend the analytic continuation into the left half-plane. Note however an interesting difference with respect to the Ising case: we cannot expect to be able to continue the free energy until we reach the negative real axis. Due to the larger angle of rotation (determined by the conformal dimensions of the operators involved), we meet the edge singularities at the points $\xi=|\xi_\mathrm{c}|\rme^{\pm i\pi41/72}$, where $|\xi_\mathrm{c}|\approx29.7$.

In the case of Ising free energy, one knows where the spinodal point and the Yang--Lee edge singularities are: the extended analiticity conjecture fills the gap in between in the $\eta$ plane (see figure 4 in \cite{JSP110:527}) so that we can state they are the same singularity. In our case, we have no `shadow domain' to make conjectures upon. The fact that the branching points we meet in the low-temperature left half-plane are precisely the edge singularities is granted by the simple dimensional analysis in the previous paragraph; what we lack is an interpretation for these points in the context of the low-temperature physics. Since they are found right where we would expect to find a spinodal singularity, except that instead of one point on the real axis we get two complex conjugate points, it is very appealing to  call them a pair of complex conjugate spinodal points.

\subsection{Free energy of $\mathcal{A}[g_1,g_2,g_4]$} \label{sec:3slice}

Once the analytic structure is revealed in the $g_3=g_4=0$ plane, we can try to explore what happens if one turns on a small perturbation in the chemical potential. Due to the fact that $h_4=3/5$ is well above the critical value 1/2 for the appearance of ultraviolet divergences, the TCS can probe only  a small region around $g_4=0$. What emerges neatly is the fact that increasing $g_4$ the Lee--Yang-like branch points get closer to the real axis. Actually, this agrees with the overall picture we presented so far. Let us see why.

The theory $\mathcal{A}_4^+$ describes the massless flow between the tricritical and the usual Ising model. From lattice viewpoint, this is related to the fact that if in the Blume--Capel model we send $\Delta\to -\infty$, then the spin-0 mode is decoupled and we recover the Ising model. Along this renormalization group flow, the conformal dimension of the magnetic field changes from 3/80 to 1/16 while the conformal dimension of the thermal perturbation evolves from 1/10 to 1/2. Hence the dimensionless ratio $\xi=g_1/|g_2|^{77/72}$ is turned into its Ising counterpart $\xi_\mathrm{IM}=g_1/|g_2|^{15/8}$. This implies that the critical values of $\xi$ identifying the position of the edge singularity are comparable, therefore we expect that along the flow generated by $g_4>0$ the value $|\xi_\mathrm{c}|\approx 29.7$ is lowered up to $|\xi_\mathrm{c}|\approx 0.0060335(7)$ that is the position of the edge singularity in the Ising model~\cite{JSP110:527}. By the same token, we expect that as $g_4\to-\infty$ the edge singularities are moved further and further from the real axis, since if we remove the values $\pm1$, no phase transition is possible.

\begin{figure}
 \centering
 \mbox{\subfigure[Low-temperature $\xi$-plane]{\epsfig{figure=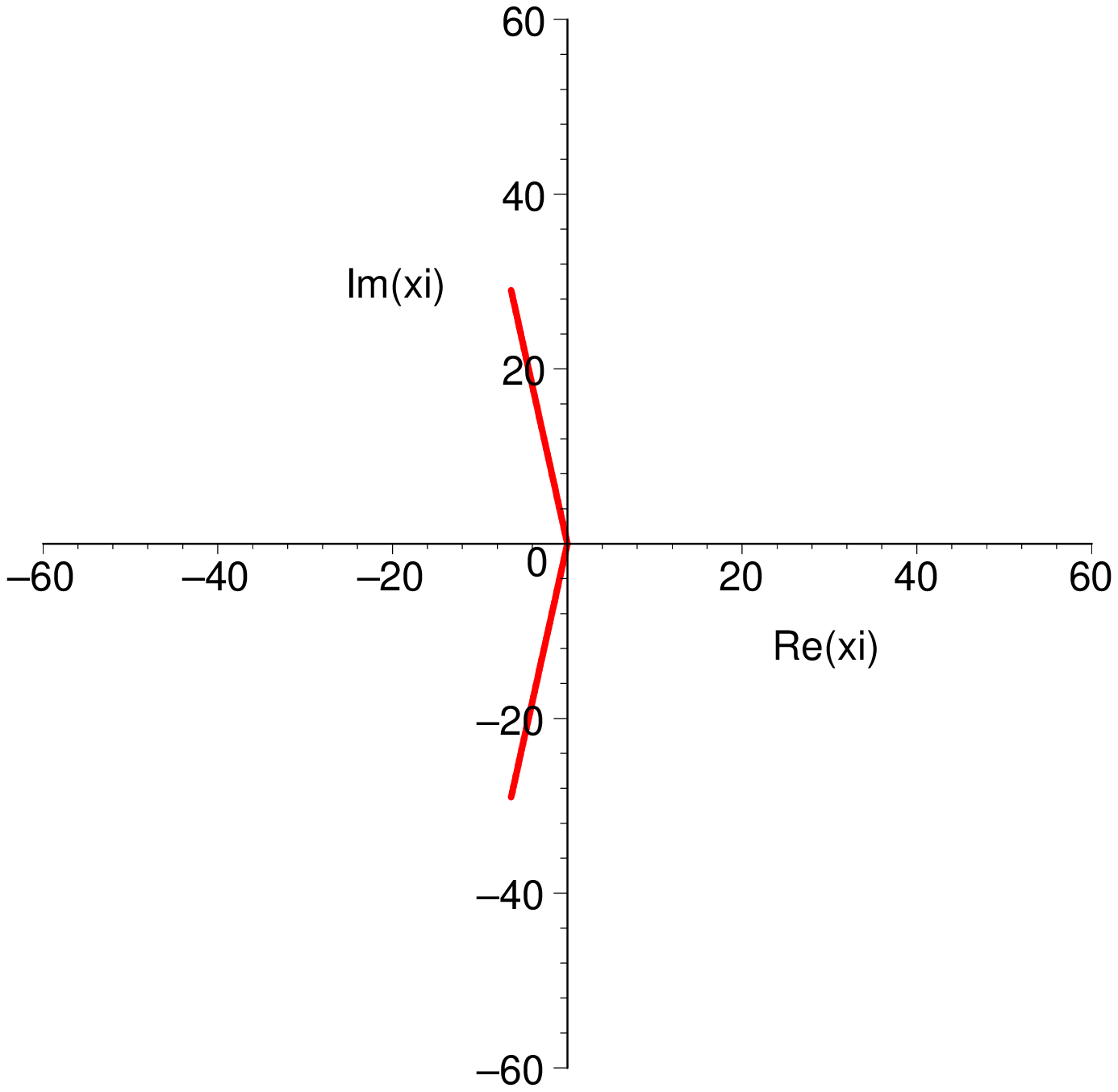, width=6cm}}
       \subfigure[High-temperature $\xi$-plane]{\epsfig{figure=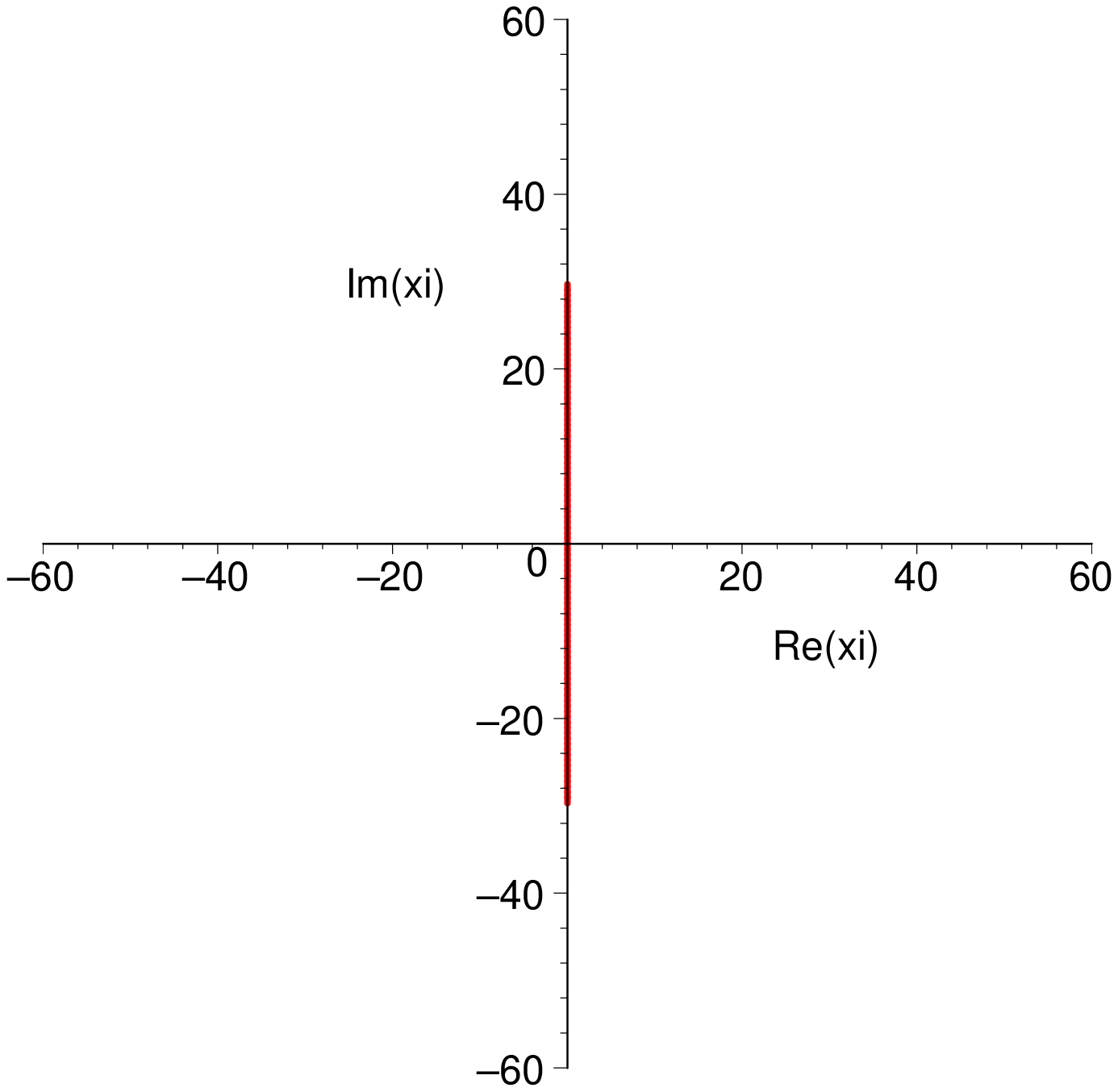, width=6cm}}}
 \mbox{\subfigure[$\eta$-plane]{\epsfig{figure=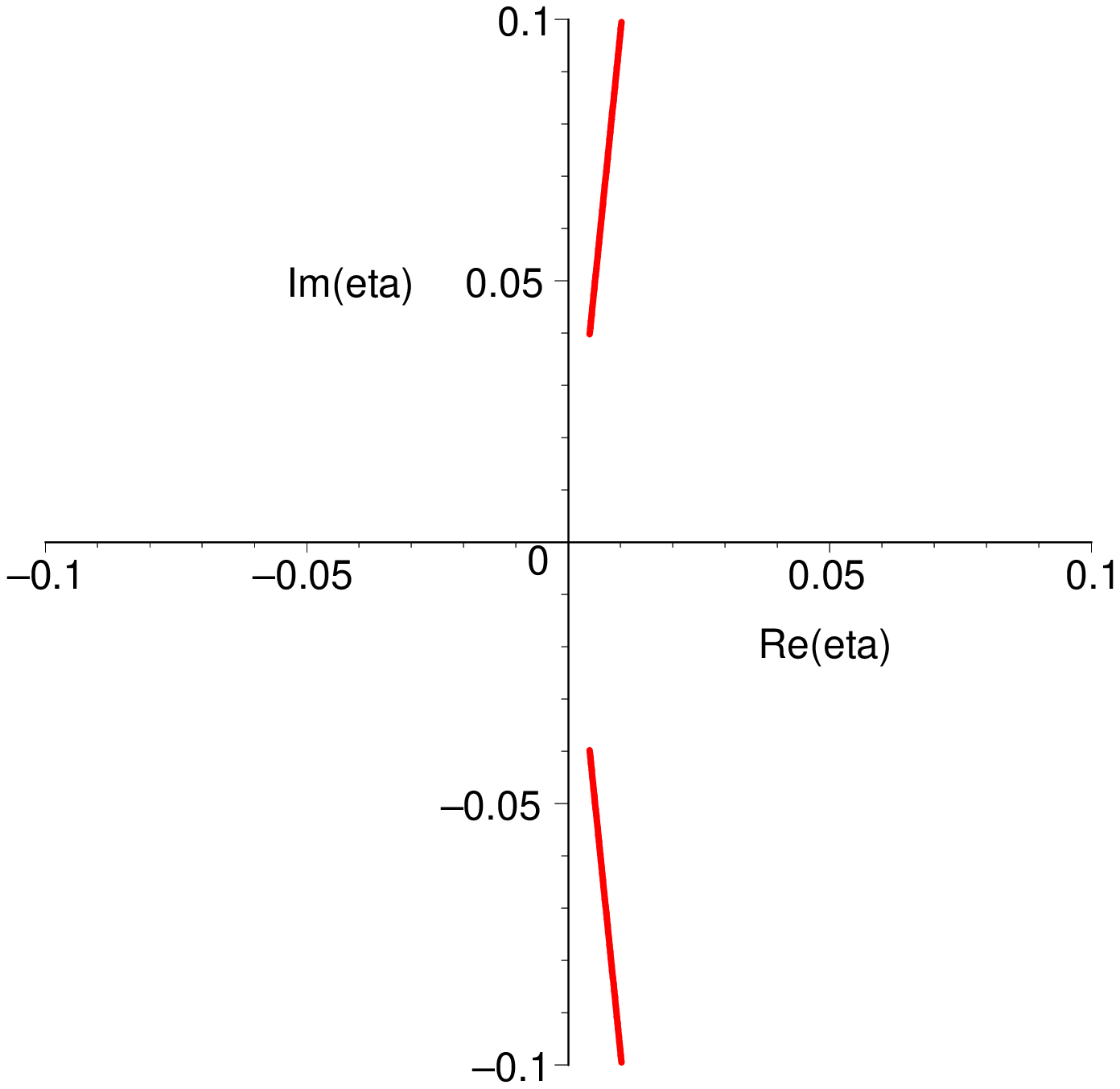,width=6cm}}}
  \caption[Alternative cuts for the theory $\mathcal{A}(g_1,g_2)$]{A different convention for the branch cuts of the theory $\mathcal{A}[g_1,g_2]$. Rotating the cuts permits to analytically continue the definition of the free energy to the left low-temperature $\xi$ half-plane, but it is impossible to reach the real negative axis as happens in Ising.} \label{fig:tagli2}
\end{figure}

Of course, we cannot claim that the observed behavior, within a range $-0.01<\zeta'<0.01$ ($\zeta'$ is defined by \eref{eq:defzeta}) is a proof of whatsoever: the best we can say is that it is a step in the expected direction.   

\subsection{Amplitudes and RG flows} \label{sec:byprod}

An extensive study about the universal amplitude ratios for the tricritical Ising model in two dimensions was performed by Fioravanti, Mussardo, and Simon~\cite{PRE63:016103}. In that paper, the truncation of conformal space is applied to the theories defined by the perturbed actions
\begin{equation} \label{eq:singpert}
  \mathcal{A}_i=\mathcal{A}_{(c=7/10)}+ g_i\int d^2x\,\varphi_i(x) \qquad
  i=1,\dots,4 \,.
\end{equation}
Since they use the eigenvectors of the truncated Hamiltonian to determine the vacuum expectation values of the primary fields~\cite{PLB411:127}, the authors of \cite{PRE63:016103} are able to evaluate the susceptibilities without introducing a second perturbation. In our study of the free energy, we computed some of these susceptibilities by means of the doubly perturbed Hamiltonians
\begin{equation} 
  \mathcal{A}_{ij}=\mathcal{A}_{(c=7/10)}+ g_i\int d^2x\, \varphi_i(x)
  +g_j\int d^2x\, \varphi_j(x) \,, 
\end{equation} 
thus obtaining as a byproduct an internal consistency check of the TCS approach. The amplitudes we computed are collected in \tref{tab:compamp} together with the results coming from integration of the correlation function and TCS applied to the theory \eref{eq:singpert}.

\begin{table}
 \caption{Amplitudes related to the measure of the free energy.} \label{tab:compamp}
 \begin{indented}
 \lineup  
 \item[]\begin{tabular}{@{}*{5}{l}}
 \br
    Amplitude & Integration$^{*}$ & ${\mathrm{TCS}^1}^*$ & $\mathrm{TCS}^2$ & Exact \cr
    \mr 
   $\Gamma_{11}^{2+}$ & 0.093(9) & 0.093(7) & 0.093(8) & \\[6pt]
   $\Gamma_{11}^{2-}$ & 0.026(2) & 0.026(7) & 0.023(7) & \\[6pt]
   $B_{12}$ & & 1.59(0) & 1.594(7) & 1.59427\dots \\[6pt]
   $B_{21}$ & & 1.35(6) & 1.33(5)  & \\[6pt]
   $B_{32}$ & & 2.3(8) & 2.5(3) & 2.45205\dots \\[6pt]
   $B_{42}$ & & 3.(4) & 3.(4) & 3.70708\dots \\[6pt]   
\br
  \end{tabular}
    \item[] $^*$ Values taken from \cite{PRE63:016103}. 
   \end{indented}
\end{table}

\begin{figure} \begin{center}
  \epsfig{figure=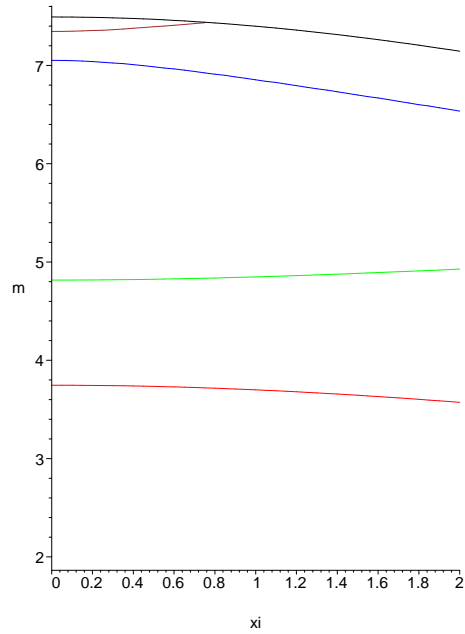, width=6cm}
  \caption[Mass spectrum evolution: from $\mathcal{A}_2^+$ to $\mathcal{A}_1$]{Mass spectrum evolution: from $\mathcal{A}_2^+$ to $\mathcal{A}_1$. The black line is the threshold $2m_1$. The four coloured lines represents the stable masses.} \label{fig:mass+21}
\end{center} \end{figure}

Another interesting possibility offered by the TCS applied on doubly perturbed conformal theories is to follow the renormalization group flow from one theory to another, as illustrated in \fref{fig:mass+21}. By turning on a magnetic field $\varphi_1$, it is possible to see the four particles under threshold of the theory $\mathcal{A}_2^+$ as they gradually become the three particles of the spectrum of $\mathcal{A}_1$.
Around the point $\xi=0.8$ the mass $m_4$ disappears as it goes above the threshold. Note that for all the levels the leading mass correction is quadratic in the magnetic field: this is not surprising, since the linear correction to the mass is related to the two-particle form factor~\cite{NPB473:469} of the field $\varphi_1$, that vanishes for symmetry reasons. Another effect of the spin-reversal symmetry is evident in the different sign that the leading correction has depending on the mass being even or odd. The ratio between the leading corrections to the masses $m_1$ and $m_2$ is a universal quantity 
\begin{equation}
  \frac{\delta m_2}{\delta m_1}=-0.685(8)
\end{equation}
which would not be easily computed in form factor approach since it involves three-particles form factors. The other two universal ratios
\begin{equation}
  \frac{\delta m_3}{\delta m_1}=6.6(4) \qquad \qquad \frac{\delta m_4}{\delta m_1}=-5.2(6) 
\end{equation}
are one order of magnitude less precise.

\section{Comments} \label{sec:comm}

In the spirit of a recent paper by Fonseca and Zamolodchikov~\cite{JSP110:527}, we have studied the analytic properties of the free energy of the tricritical Ising model. This minimal model of conformal field theory is characterized by four relevant primary operators, that can be interpreted as a leading and a subleading magnetic fields, a thermal and a chemical potential perturbations.

 The main technical tool of our research is the truncation of conformal space, an approximate technique that gives access to the spectrum of the perturbed conformal field theory put on a cylinder. Such a numerical method permits to get nonperturbative data about nonintegrable field theories. The main limitation of TCS is its bad performance when dealing with perturbations originated by relevant fields whose conformal dimension is near to or greater than 1/2. This failure is inherent to the approach, so we cannot expect to be able to probe by means of it the behavior of the free energy far from the plane $g_3=g_4=0$. 

Such plane can indeed be studied, and the results of this inspection bring some expected facts and some surprising ones. The high-temperature sector exhibits a pair of Lee--Yang-like edge singularities on the imaginary axis of the magnetic field, like the simpler Ising model. The nonunitary minimal model related to the critical process of accumulation of the zeros of the partition function could be $\mathcal{M}(7,2)$, as claimed by von Gehlen~\cite{IJMPB8:3507}, but we failed in producing some euristhic theoretical argument in favor of his numerical finite-size-scaling results. The direct computation of the effective central charge associated to the renormalization group flow is within the possibility of TCS, but requires an heavy increase in computational cost that we did not pursue yet.

On the side of the surprising results, there is an interesting difference with respect to the Ising prototype. In the low-temperature regime, due to the different conformal dimensions of the operators involved, it is not possible, starting from the positive real magnetic field axis, to analytically continue the definition of the free energy until reaching the negative real axis. It seems appealing to identify the two unexpected branching points that appear in the left half-plane with a pair of complex conjugate spinodal singularities. This identification, much in the spirit of the extended analyticity conjecture proposed by Fonseca and Zamolodchikov, opens a problem of physical interpretation.

We tried to enlarge the analysis by including also the coupling $g_4$. An euristhic argument, based on the physical interpretation of $g_4$ and on the existence of a massless renormalization group flow from TIM to Ising field theory along the $\varphi_4$ perturbation, suggests what may be the extension of our results to the three dimensional space identified by $g_3=0$. As far as this conjecture can be numerically tested by TCS -- that is not very far, to be honest -- it seems to hold. By increasing the level of truncation one can hope to probe also the $g_3$ direction, but a huge computational effort will be required.

As a byproduct of our work on the analytic properties, we could compute some susceptibilities and vacuum expectation values already evaluated by means of a slightly different use of TCS, thus obtaining an internal consistency check. Moreover, we studied the evolution of the mass spectrum of the theory deformed by the leading energy perturbation (in the high-temperature regime) when a small magnetic field is introduced. The universal ratios between mass corrections have been determined for the first time.    

\ack
AM is indebted to G Delfino, G Feverati and P Grinza for many useful discussions. 


\section*{References}
\bibliographystyle{iopart-num}
\bibliography{tesi}

\end{document}